\documentclass[iop,apj]{emulateapj}

\usepackage{graphicx}
\usepackage{epstopdf}
\usepackage{amsmath}
\usepackage{rotating}

\font\rmsmall=cmr8
\newcommand\litl{\rm\scriptscriptstyle}
\newcommand{\expo}[1]{\ensuremath{10^{#1}}}
\newcommand{\nexpo}[2]{\ensuremath{{#1}\times10^{#2}}}

\newcommand{\myemail}{ingalls\@ipac.caltech.edu}

\newcommand{\etal}{{\rm et al.},}

\newcommand{\ie}{{\rm i.e.},}
\newcommand{\kms}{\ensuremath{\,\km\s^{-1}}}

\newcommand{\wmsr}{\ensuremath{\,{\rm  W}\,{\rm m}^{-2}\persr}}

\newcommand{\persr}{\ensuremath{\,{\rm sr}^{-1}}}
\newcommand{\cmtwo}{\ensuremath{\,{\rm cm}^{-2}}}

\newcommand\cmthree{\ensuremath{\,{\rm cm}^{-3}}}
\newcommand\K{\ensuremath{\,{\rm K}}}

\newcommand\km{\ensuremath{\,{\rm  km}}}

\newcommand\s{\ensuremath{\,{\rm  s}}}

\newcommand\yr{\ensuremath{\,{\rm  yr}}} 
\newcommand\pc{\ensuremath{\,{\rm  pc}}}

\newcommand\mjysr{\ensuremath{\,{\rm MJy}\,{\rm sr}^{-1}}}

\newcommand\htwo{\ensuremath{{\rm H}_2}}

\newcommand{\carbon}[1]{\ensuremath{^{#1}{\rm C}}}

\newcommand\co  {\carbon{12}{\rm O}}

\newcommand{\cii}{{\sc C~ii}\/}
\newcommand{\ci}{{\sc C~i}\/}
\newcommand{\oi}{{\sc O~i}\/}
\newcommand{\hi}{{\sc H~i}\/}
\newcommand{\hii}{{\sc H~ii}\/}
\newcommand\av{\ensuremath{A_{\rm v}}}

\newcommand\nnh{\ensuremath{{\rm N}_{\litl H}}}
\newcommand\nh{\ensuremath{{{\rm n}_{\litl H}}}}

\newcommand\trot{\ensuremath{T_{r}}}

\newcommand{\Spitzer}{\hbox{\sl Spitzer\/}}
\newcommand{\pdeg}{\ensuremath{{}^{\circ}\mskip-8mu.\,}} 

\slugcomment{Accepted for publication in ApJ 2011 October 19}

\shorttitle{\htwo\ Rotational Emission in Translucent Clouds}
\shortauthors{Ingalls et al.}

\begin{document}


\title{Spitzer IRS Detection of Molecular Hydrogen\\ Rotational Emission towards Translucent Clouds}


\author{James G. Ingalls\altaffilmark{1}, T. M. Bania\altaffilmark{2}, F. Boulanger\altaffilmark{3}, B. T. Draine\altaffilmark{4}, E. Falgarone\altaffilmark{5}, and P. Hily-Blant\altaffilmark{6}}
\altaffiltext{1}{Spitzer Space Telescope Science Center, California Institute of Technology, 1200 E California Blvd, Mail Stop 220-6, Pasadena, CA 91125; ingalls@ipac.caltech.edu}
\altaffiltext{2}{Institute for Astrophysical Research, Boston University, 725 Commonwealth Avenue,
Boston, MA 02215; bania@bu.edu}
\altaffiltext{3}{Institut d'Astrophysique Spatiale, Universit\'e Paris Sud, Bat. 121, 91405 Orsay Cedex, France; francois.boulanger@ias.u-psud.fr}
\altaffiltext{4}{Princeton University Observatory, Peyton Hall, Princeton, NJ 08544; draine@astro.princeton.edu}
\altaffiltext{5}{Laboratoire de Radio-Astronomie, LERMA, Ecole Normale Sup\'erieure, 24 rue Lhomond, 75231 Paris Cedex 05, France; edith.falgarone@lra.ens.fr}
\altaffiltext{6}{LAOG, CNRS UMR 5571, UniversitŽ Joseph Fourier, BP53, 38041 Grenoble, France; pierre.hilyblant@obs.ujf-grenoble.fr}

\begin{abstract}
Using the Infrared Spectrograph on board the \Spitzer\ Space Telescope, we have detected emission in the S(0), S(1), and S(2) pure-rotational ($v=0-0$) transitions of molecular hydrogen (\htwo) towards 6 positions in two translucent high Galactic latitude clouds, DCld 300.2$-$16.9 and LDN 1780.  The detection of these lines raises important questions regarding the physical conditions inside low-extinction clouds that are far from ultraviolet radiation sources.  The ratio between the S(2) flux and the flux from PAHs at 7.9 microns averages 0.007 for these 6 positions. This is a factor of about 4 higher than the same ratio measured towards the central regions of non-active Galaxies in the \Spitzer\ Infrared Nearby Galaxies Survey (SINGS). Thus the environment of these translucent clouds is more efficient at producing rotationally excited \htwo\ per PAH-exciting photon than the disks of entire galaxies.  Excitation analysis finds that the S(1) and S(2) emitting regions are warm ($T\gtrsim 300\K$), but comprise no more than 2\% of the gas mass.  We find that UV photons cannot be the sole source of excitation in these regions and suggest mechanical heating via shocks or turbulent dissipation as the dominant cause of the emission. The clouds are located on the outskirts of the Scorpius-Centaurus OB association and may be dissipating recent bursts of mechanical energy input from supernova explosions.  We suggest that pockets of warm gas in diffuse or translucent clouds, integrated over the disks of galaxies, may represent a major source of all non-active galaxy \htwo\ emission.
\end{abstract}




\section{Introduction}
The kinetic temperature of all cold ($T < 100$\,K) diffuse interstellar clouds is the result of a balance between heating and cooling processes.  The \hi\ and \htwo\ receive kinetic energy from electrons ejected from dust grains and polycyclic aromatic hydrocarbons (PAHs) by interstellar far-ultraviolet (FUV; 6\,eV$< h\nu< 13.6$\,eV) photons, and transfer that energy to all other gas phase constituents.  Clouds are cooled when inelastic collisions lead to spontaneous emission of spectral line radiation.  Despite its being the most abundant constituent of molecular clouds, the role of \htwo\ in directly cooling the diffuse ISM via its own line emission is uncertain.  \citet{spi49} was the first to estimate the significance of the pure-rotational ($v=0-0$) transitions of \htwo\ in cooling interstellar matter. Since that time, it has been realized that other species (\cii\ in atomic gas and CO in UV-shielded molecular gas) were much more efficient at converting kinetic energy to line emission, despite their much lower abundances ($\sim \expo{-4}$) relative to hydrogen \citep{wol03,hol91}.

Because of the small moment of inertia of \htwo, its rotational energy levels  are difficult to excite in the cold ISM.  The lowest-lying transitions are the electric quadrupole pure-rotational lines.  Because \htwo\ lacks an electric dipole moment, the $J=1$ level is unable to radiate, so the first radiating level is $J=2$. The first three rotational transitions are thus S(0) ($J=2-0$) at 28.2\micron, S(1) ($J=3-1$) at 17.0\micron, and S(2) ($J=4-2$) at 12.3\micron.  All were observable with the \Spitzer\ InfraRed Spectrograph (IRS).  The lowest transition S(0) has $h\nu/k = 510\,$K.  The S(2) transition occurs from a rotational level ($J=4$) that is about 1700\,K above the ground state. 
\begin{deluxetable*}{lccrrrrrrrrr}

\tabletypesize{\scriptsize}
\tablewidth{7.2in}
\tablecaption{Observed \htwo\  and Dust Surface Brightness}
\tablehead{
~&~&~&\multicolumn{7}{c}{\htwo\ Pure-Rotational Surface Brightness\tablenotemark{a}}&~&~\\
~&~&~&\multicolumn{7}{c}{\hrulefill}&~&~\\
\multicolumn{1}{c}{Position} & \multicolumn{1}{c}{RA (2000)} & \multicolumn{1}{c}{~DEC (2000)}& \multicolumn{1}{c}{S(0)} & \multicolumn{1}{c}{S(1)}  & \multicolumn{1}{c}{S(2)} & \multicolumn{1}{c}{S(3)} & \multicolumn{1}{c}{S(4)}  & \multicolumn{1}{c}{S(5)}  & \multicolumn{1}{c}{S(6)}&\multicolumn{1}{c}{$I_{\litl 7.9,PAH}$\tablenotemark{b}}&\multicolumn{1}{c}{TIR\tablenotemark{c}}\\
& \multicolumn{1}{r}{(hh:mm:ss.s)} & \multicolumn{1}{r}{(dd:mm:ss)}&\multicolumn{7}{c}{(\expo{-10}\wmsr)}&\multicolumn{1}{c}{(\expo{-7})}&\multicolumn{1}{c}{(\expo{-6})}
}
\startdata
\dataset[ADS/sa.spitzer\#0022715136]{DCld 300.2-16.9 (C)} & 11:48:24.4~ & --79:18:00~ &36.74 & 13.5 & 12.8 & ~$<4.0$\tablenotemark{d} & $<6.8$ & $<4.7$ & $<8.0$& 2.436 &1.64\\
~&~&~&$\pm$0.16&$\pm$0.4&$\pm$0.5&&&&&$\pm$0.007&$\pm$0.16\\

\dataset[ADS/sa.spitzer\#0022715136]{DCld 300.2-16.9 (B)} & 11:52:08.3~ & --79:09:33~ & 29.43&13.5&8.9&$<3.8$ & $<5.8$ & $<5.3$ & $<11.2$& 2.745 &3.4$\phantom{0}$\\
~&~&~&$\pm$0.16&$\pm$0.4&$\pm$0.4&&&&&$\pm$0.015&$\pm$0.3$\phantom{2}$\\

\dataset[ADS/sa.spitzer\#0014030848]{DCld 300.2-16.9 (A)} & 11:54:23.1~ & --79:31:42~ & 24.03 & 9.4& 7.7& $<3.4$ & $<4.0$ & $<4.3$ & $<5.4$& 2.137&1.51\\
~&~&~&$\pm$0.15&$\pm$0.4&$\pm$0.5&&&&&$\pm$0.018&$\pm$0.15\\

\dataset[ADS/sa.spitzer\#0022715136]{DCld 300.2-16.9 (D)} &  11:55:33.8~ & --79:20:54~ & 26.27 & 9.7 & 9.0 & $<4.2$ & $<5.1$ & $<5.0$ & $<5.4$&1.714 &0.83\\
~&~&~&$\pm$0.14&$\pm$0.4&$\pm$0.5&&&&&$\pm$0.005&$\pm$ 0.08\\

\dataset[ADS/sa.spitzer\#0022715392]{LDN 1780 (7)} & 15:39:23.0~ & --07:10:05~ & \nodata & \nodata & 6.9 & $<5.4$ & $<3.2$ & $<1.5$ & $<3.2$&0.742 &1.47 \\
~&~&~& & &$\pm$0.6&&&&&$\pm$0.003&$\pm$0.14\\

\dataset[ADS/sa.spitzer\#0022715648]{LDN 1780 (3)} & 15:40:34.2~ & --07:13:14~ & \nodata& \nodata & 20.9 & $<4.6$ & $<5.4$ & $<9.5$ & $<7.6$&1.533&2.7\\
~&~&~&&&$\pm$0.6&&&&&$\pm$0.002&$\pm$0.3\\

\enddata
\tablenotetext{a}{Integrated \htwo\ surface brightness from spectrum fits, in units of \expo{-10}\wmsr.}
\tablenotetext{b}{$I_{\litl 7.9,PAH}\equiv \langle\nu I_\nu\rangle_{7.9\mu{\rm m}}$
measured over the {\sl Spitzer} IRAC 7.9$\mu$m band.   Units are $\expo{-7}\wmsr$.}
\tablenotetext{c}{Estimated total infrared surface brightness, in units of \expo{-6}\wmsr.}
\tablenotetext{d}{Upper limits are 3$\sigma$, as described in the text.}
\label{srctable}
\end{deluxetable*}

Knowledge of the distribution of physical conditions along a given line of sight is required for the proper interpretation of any observation of \htwo\ lines.  This is especially true for path lengths spanning many kpc such as the disks of galaxies.  Since these quadrupole transitions require high temperatures to excite by collisions, the presence of \htwo\ lines in the ISM of galaxies is usually thought to be the result of either collisional excitation in warm \htwo\ or UV pumping.  Collisional excitation is expected to dominate in shock waves, while UV pumping dominates the excitation in photodissociation regions (PDRs) \citep[eg.,][]{shu78,ber00,mey01}.  In other cases, however, the source of \htwo\ excitation is not so clear.  In the recent \Spitzer\ Infrared Nearby Galaxies Survey \citep[SINGS;][]{ken03} measurements of \htwo\ in normal galaxies, \citet{rou07} concluded that FUV-rich photodissociation regions (PDRs) at the interfaces between bright \hii\ regions and molecular clouds were the most likely source of the emission.  The authors did note that some of the \htwo\ flux could originate in diffuse gas illuminated by the ambient FUV field ($G_0\sim 1$, in units of the average interstellar radiation field near the Sun), which is in fact the source of most of the IR emission in the same sample \citep{dra07b}.  But their paper did not favor this interpretation of their data.

A growing body of analysis finds that the \htwo\ emission from the disks of galaxies {\it does} originate partly in diffuse material.  Using FUSE measurements of FUV absorption by \htwo\ towards late B stars behind the Chamaeleon complex, \citet{gry02} found that the rotational levels could not be populated by only FUV photons, and required collisional excitation within anomalously warm regions.  Indeed, further study of the cirrus cloud in front of these stars \citep{neh08a} has determined that the observed CH$^+$ absorption also depends on the existence of warm pockets of gas, since the formation of CH$^+$ molecules needs an activation energy of $\Delta E/k = 4640\,$K.  The first detection of pure-rotational \htwo\ emission from diffuse Galactic regions was the \citet{fal05} ISO SWS measurement of the S(0) to S(3) transitions along a direction in the mid-plane of the Milky Way (MW: $l,b$=26\pdeg46,0\pdeg09)  As concluded from the absorption result of \citet{gry02}, the MW \htwo\ emission turns out to be {\it too bright} to be caused only by FUV excitation.   

In this paper, we give additional evidence that low extinction, low-UV environments can excite \htwo\ molecules.  In the following section we describe recent ultra-sensitive \Spitzer\ Space Telescope observations of the 5--15\micron\ (5--38\micron\ for four postions) spectrum towards translucent clouds at high Galactic latitude ($|b|\gtrsim 15\deg$) associated with infrared cirrus.  In \S3 we show detections of \htwo\ S(0) and S(1) emission towards 4 positions and S(2) emission towards 6 positions.  We estimate the physical conditions in the regions producing \htwo\ line radiation.  We compute ratios of \htwo\ flux to PAH feature strength and compare with the measurements of the same quantities made as part of the \Spitzer\ Legacy SINGS survey of non-active galaxies.  In \S4 we review possible sources of excitation of the \htwo\ molecules, and discuss the implications for the study of \htwo\ emission from the disks of galaxies. In \S5 we make our concluding remarks.

\section{Observations}
\begin{deluxetable*}{lccrrrrrc}

\tabletypesize{\scriptsize}
\tablewidth{6.1in}
\tablecaption{Positions with \htwo\ Upper Limits and Dust Surface Brightness}
\tablehead{
~&~&~&\multicolumn{5}{c}{\htwo\ Pure-Rotational Upper Limits\tablenotemark{a}}&~\\
~&~&~&\multicolumn{5}{c}{\hrulefill}&~\\
\multicolumn{1}{c}{Position} & \multicolumn{1}{c}{RA (2000)} & \multicolumn{1}{c}{~DEC (2000)}& \multicolumn{1}{c}{S(2)} & \multicolumn{1}{c}{S(3)} & \multicolumn{1}{c}{S(4)}  & \multicolumn{1}{c}{S(5)}  & \multicolumn{1}{c}{S(6)}&\multicolumn{1}{c}{$I_{\litl 7.9,PAH}$\tablenotemark{b}}\\
& \multicolumn{1}{r}{(hh:mm:ss.s)} & \multicolumn{1}{r}{(dd:mm:ss)}&\multicolumn{5}{c}{(\expo{-10}\wmsr)}&\multicolumn{1}{c}{(\expo{-7})}
}
\startdata
\dataset[ADS/sa.spitzer\#14031104]{MBM 12 (01)} & 02:56:09.6 &+19:26:51  & 39.3 & 60.7 & 90.4 & 54.7 & 85.6 & 1.264$\pm$0.018\\
\dataset[ADS/sa.spitzer\#0022714112]{MBM 12 (A; -509.1,+636.4)}\tablenotemark{c} &  02:56:19.9 &+19:26:04 & 2.6 & 4.1 & 2.6 & 8.4 & 3.7 & 0.573$\pm$0.007\\
\dataset[ADS/sa.spitzer\#0022714112]{MBM 12 (A; -381.8,+509.1)} & " & " & 2.2 & 4.7 & 2.5 & 8.5 & 5.8 & 0.681$\pm$0.022\\\dataset[ADS/sa.spitzer\#0022714112]{MBM 12 (A; -254.6,+381.8)} & " & " & 11.6 & 40.2 & 28.2 & 15.6 & 23.7 & 1.09$\pm$0.03\\
\dataset[ADS/sa.spitzer\#22714624]{MBM 12 (A; -254.6,+381.8)} & " & " & 11.6 & 40.2 & 28.2 & 15.6 & 23.7 & 1.09$\pm$0.03\\
\dataset[ADS/sa.spitzer\#22714624]{MBM 12 (A; -127.3,+254.6)} & " & " & 15.6 & 78.0 & 35.7 & 23.6 & 33.7 & 0.88$\pm$0.05\\
\dataset[ADS/sa.spitzer\#22714624]{MBM 12 (A; 0,+127.3)} & " & " & 12.4 & 52.6 & 24.9 & 24.6 & 23.6 & 0.62$\pm$0.03\\
\dataset[ADS/sa.spitzer\#22714624]{MBM 12 (A; +127.3,-254.6)} & " & " & 14.0 & 102.0 & 32.3 & 22.8 & 28.8 & 0.78$\pm$0.24\\
\dataset[ADS/sa.spitzer\#22714624]{MBM 12 (A; +254.6,-381.8)} & " & " & 9.0 & 26.8 & 17.5 & 16.2 & 18.6 & 0.962$\pm$0.022\\
\dataset[ADS/sa.spitzer\#0022714112]{MBM 12 (A; +381.8,-509.1)} & " & " & 2.9 & 4.1 & 3.8 & 7.0 & 3.6 & 0.634$\pm$0.015\\
\dataset[ADS/sa.spitzer\#0022714112]{MBM 12 (A; +509.1,-636.4)} & " & " & 2.5 & 5.4 & 2.7 & 9.7 & 4.9 & 0.570$\pm$0.006\\
\dataset[ADS/sa.spitzer\#22713600]{MBM 12 (10)} & 02:56:40.4 & +19:24:20 & 24.2 & 87.3 & 72.4 & 41.9 & 57.6 & 1.15$\pm$0.03\\
\dataset[ADS/sa.spitzer\#14031360]{MBM 12 (01 off)} & 02:57:07.9 & +19:16:17 & 14.8 & 38.2 & 38.0 & 22.0 & 29.9 & 1.039$\pm$0.022\\
\dataset[ADS/sa.spitzer\#14031616]{MBM 12 (12)} & 02:57:27.2 &+20:02:45  & 33.7 & 51.3 & 79.9 & 56.8 & 61.7 & 1.866$\pm$0.019\\
\dataset[ADS/sa.spitzer\#14031872]{MBM 12 (12 off)} & 02:58:39.0 & +20:14:16 & 45.0 & 69.4 & 106.7 & 62.5 & 85.0 & 1.472$\pm$0.017\\
\dataset[ADS/sa.spitzer\#14032128]{MBM 28 (A)}\tablenotemark{d}  & 09:29:09.2 & +70:31:00 & 35.5 & 55.0 & \nodata & \nodata & \nodata & \nodata\\
\dataset[ADS/sa.spitzer\#22713344]{DCld 300.2-16.9 (E)} & 11:46:18.9 & --78:46:33 & 1.3 & 2.5 & 3.1 & 7.2 & 5.6 & 0.562$\pm$0.011\\
\dataset[ADS/sa.spitzer\#22712832]{LDN 183 (W1; -540,0)} & 15:52:54.5 & --02:52:24 & 2.7 & 7.9 & 6.7 & 6.1 & 7.6 & 1.341$\pm$0.018\\
\dataset[ADS/sa.spitzer\#22712832]{LDN 183 (W1; -360,0)} & " & " & 2.9 & 6.6 & 6.9 & 3.8 & 4.7 & 1.527$\pm$0.015\\
\dataset[ADS/sa.spitzer\#22712832]{LDN 183 (W1; -180,0)} & " & " & 3.8 & 9.2 & 10.0 & 6.6 & 7.6 & 1.675$\pm$0.017\\
\dataset[ADS/sa.spitzer\#22712832]{LDN 183 (W1; 0,0)} & " & " & 3.2 & 9.2 & 11.2 & 6.5 & 5.0 & 1.329$\pm$0.013\\
\dataset[ADS/sa.spitzer\#22713344]{LDN 183 (W1; +180,0)} & " & " & 2.7 & 4.0 & 7.3 & 13.9 & 13.2 & 1.033$\pm$0.009\\
\dataset[ADS/sa.spitzer\#22713344]{LDN 183 (W1; +360,0)} & " & " & 2.7 & 4.8 & 6.9 & 12.9 & 13.2 & 0.960$\pm$0.018\\
\dataset[ADS/sa.spitzer\#22713344]{LDN 183 (W1; +540,0)} & " & " & 2.8 & 5.0 & 6.7 & 13.3 & 14.3 & 0.902$\pm$0.009\\
\dataset[ADS/sa.spitzer\#14032384]{Stark 4 (S1; -480,0)} &  23:45:07.3 & --71:42:47 & 5.2 & 9.1 & 13.7 & 12.3 & 16.2 & $<0.340$\tablenotemark{e}\\
\dataset[ADS/sa.spitzer\#14032640]{Stark 4 (S1; -240,0)} & " & " & 3.3 & 5.4 & 7.4 & 6.1 & 8.4 & 0.113$\pm$0.016\\
\dataset[ADS/sa.spitzer\#14032896]{Stark 4 (S1; 0,0)} & " & " & 2.1 & 3.4 & 5.3 & 5.2 & 6.4 & 0.206$\pm$0.019\\
\dataset[ADS/sa.spitzer\#14033152]{Stark 4 (S1; +240,0)} & " & " & 3.4 & 4.7 & 8.0 & 5.3 & 5.3 & 0.178$\pm$0.006\\
\dataset[ADS/sa.spitzer\#14033408]{Stark 4 (S1; +480,0)} & " & " & 7.7 & 10.4 & 15.5 & 11.1 & 14.0 & 0.299$\pm$0.012\\
\enddata
\tablenotetext{a}{Integrated \htwo\ surface brightness upper limits (3$\sigma$) from spectrum fits, in units of \expo{-10}\wmsr.}
\tablenotetext{b}{$I_{\litl 7.9,PAH}\equiv \langle\nu I_\nu\rangle_{7.9\mu{\rm m}}$
measured over the {\sl Spitzer} IRAC 7.9$\mu$m band.  Units are $\expo{-7}\wmsr$.}
\tablenotetext{c}{Offset from AOR center given as ($\Delta$RA,$\Delta$DEC), in arcseconds. }
\tablenotetext{d}{The SL2 portion of the spectrum for MBM 28 (A) was corrupted by stray light and did not yield measurements for S(3) through S(6), nor for $I_{\litl 7.9,PAH}$.}
\tablenotetext{e}{Upper limit to $I_{\litl 7.9,PAH}$ is 3 times the RMS in the band.}
\label{nondetec_table}
\end{deluxetable*}

\begin{figure*}
\plotone{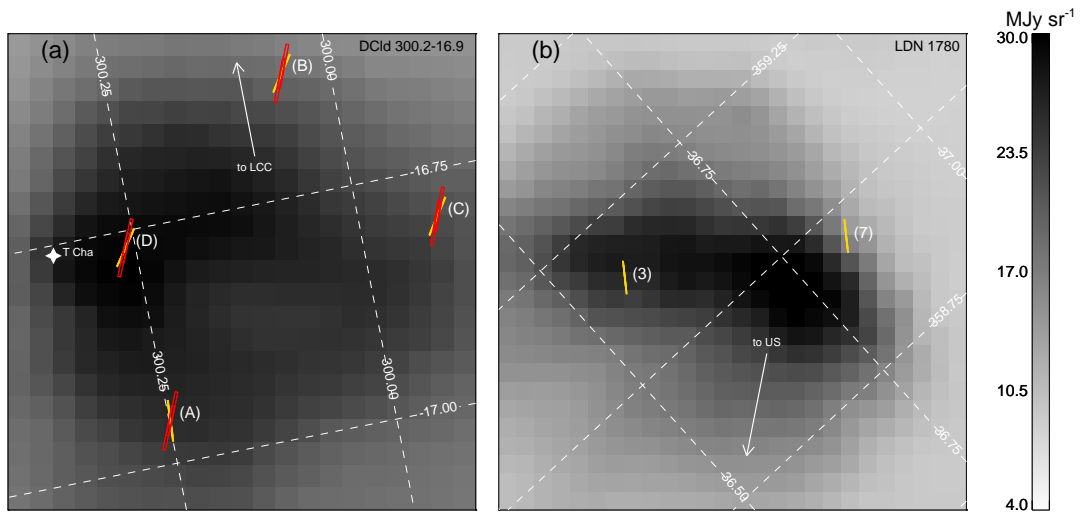}

\caption{{\sl IRAS} Sky Survey Atlas images of the 100\micron\ emission towards ({\it a}) DCld 300.2--16.9 and ({\it b}) LDN 1780.  Superposed are images of the SL ({\it yellow)} and LL ({\it red}) slit apertures used for our measurements.  The direction towards the Lower Centaurus Crux and Upper Scorpius OB associations are shown by arrows on the left and right panels, respectively.  In these images, celestial North is up and Galactic coordinates are indicated by a dashed grid. 
\label{slit_geom}}
\end{figure*}

\subsection{{\sl Spitzer} Spectroscopic Measurements}

As part of a study of the heating and cooling of high Galactic latitude translucent clouds (HLCs), we spectrally imaged 34 positions in six HLCs at 5.2--14.5\micron\ using the \Spitzer\ Space Telescope \citep{wer04} Infrared Spectrograph (IRS) \citep{hou04}.  The data were taken under General Observer programs 20093 and 40207.  We detected six positions in two of the clouds, DCld 300.2--16.9 and LDN 1780 ($l,b\approx 358\pdeg9,+36\pdeg9$), in \htwo\ S(2) pure-rotational emission at 12.3\micron.  The data for position A of cloud DCld 300.2--16.9 \dataset[ADS/sa.spitzer\#0014030848]{({\sl Spitzer} Request Key [SRK] 0014030848)} were obtained on 2005 July 9; those for positions B, C, and D \dataset[ADS/sa.spitzer\#0022715136]{(SRK 0022715136)} were obtained on 2007 July 24; and those for LDN 1780 positions 3 \dataset[ADS/sa.spitzer\#0022715648]{(SRK 0022715648)} and 7 \dataset[ADS/sa.spitzer\#0022715392]{(SRK 0022715392)} were obtained on 2008 March 26.  In addition, we imaged the 14--38\micron\ spectrum towards four positions in DCld 300.2--16.9 under Director's Discretionary Time program 491.  These measurements \dataset[ADS/sa.spitzer\#0028315392]{(SRK 0028315392)} were obtained on 2009 April 16, and yielded \htwo\ S(0) 28.2\micron\ and S(1) 17.0\micron\ detections towards all positions.  The coordinates of the \htwo-detected positions are given in Table \ref{srctable}.  Those for the undetected positions are provided in Table \ref{nondetec_table}.  Although we present some of the dust measurements in this paper, a full analysis of the interstellar dust emission in the six-cloud sample is the subject of a future paper (Ingalls et al, in preparation).  We describe in this section the method by which we measured the S(0), S(1), and S(2) emission.

Figure \ref{slit_geom} shows the orientations and areas of coverage of the IRS subslits that we used to produce spectra for the four positions in DCld 300.2--16.9 and the two positions in LDN 1780. The slit apertures are shown superimposed on 100\micron\ emission maps from the {\sl IRAS} Sky Survey Atlas (ISSA).  The 5.2--14.5\micron\ measurements of both clouds were made using the IRS short wavelength low resolution (SL) module.  Each target was observed in standard staring mode for 8 cycles of 60 seconds each.  Under staring mode, for each data collection cycle the SL2 subslit (5.2--7.7\micron), and then the SL1 subslit (7.4--14.5\micron), are each placed on the target position at two offset ``nod'' positions, 1/3 and 2/3 along the long axis of the $57\arcsec\times 3.7\arcsec$ subslit.  When one subslit observed the target, the other subslit observed a region of adjacent sky.  The two subslits are separated by 22$\arcsec$.  [For a complete description of the SL geometry, see the IRS Instrument Handbook \citep{ssc09}.]  We spent a total of about 32 minutes imaging a $234\arcsec\times 3.7\arcsec$ region around each target.  Since each subslit observes a different region of adjacent sky, the area of overlap between the two subslits was only $95\arcsec\times 3.7\arcsec$.  Nevertheless in what follows we combine all observations in the $234\arcsec$ window covered by either SL1 or SL2 (yellow in Fig. \ref{slit_geom}).  The results do not change if we limited the measurements to within the $95\arcsec$ region of overlap, although the noise increases noticeably.  

The 14-38\micron\ measurements towards 4 positions in DCld 300.2--16.9 were made using the IRS long wavelength low resolution (LL) module, whose slit covers a larger region of the sky than SL.  Each of the four targets (plus the background position---see below) was observed in staring mode for 24 cycles of 30 seconds each.  As for the SL observations, the short (LL1 subslit; 14.0-21.3\micron) and long (LL2 subslit; 19.5-38.0\micron) wavelength portions of the spectra were imaged on different portions of the sky.  In the case of LL, however, we only combined the data for the two subslit positions that overlapped on the sky, yielding about 24 minutes per position and covering a $280\arcsec\times 10.6\arcsec$ region (red on Fig. \ref{slit_geom} {\it a}).  To increase the likelihood of scheduling these measurements, we did not constrain the observation dates, and therefore the SL slit position angles were not replicated exactly by the LL measurements.  This resulted in SL and LL effective apertures that were well-matched in size, but only overlapped a relatively small portion of sky, near the centers of the imaged regions (see Fig. \ref{slit_geom}).  In other words, we have chosen to minimize random noise (maximize integration time per area of sky) at the expense of systematic error due to mismatched position angle, assuming that the emission is not a strong function of position.  
\begin{sidewaysfigure*}
\vspace{4in}
\includegraphics[keepaspectratio=true,scale=0.75]{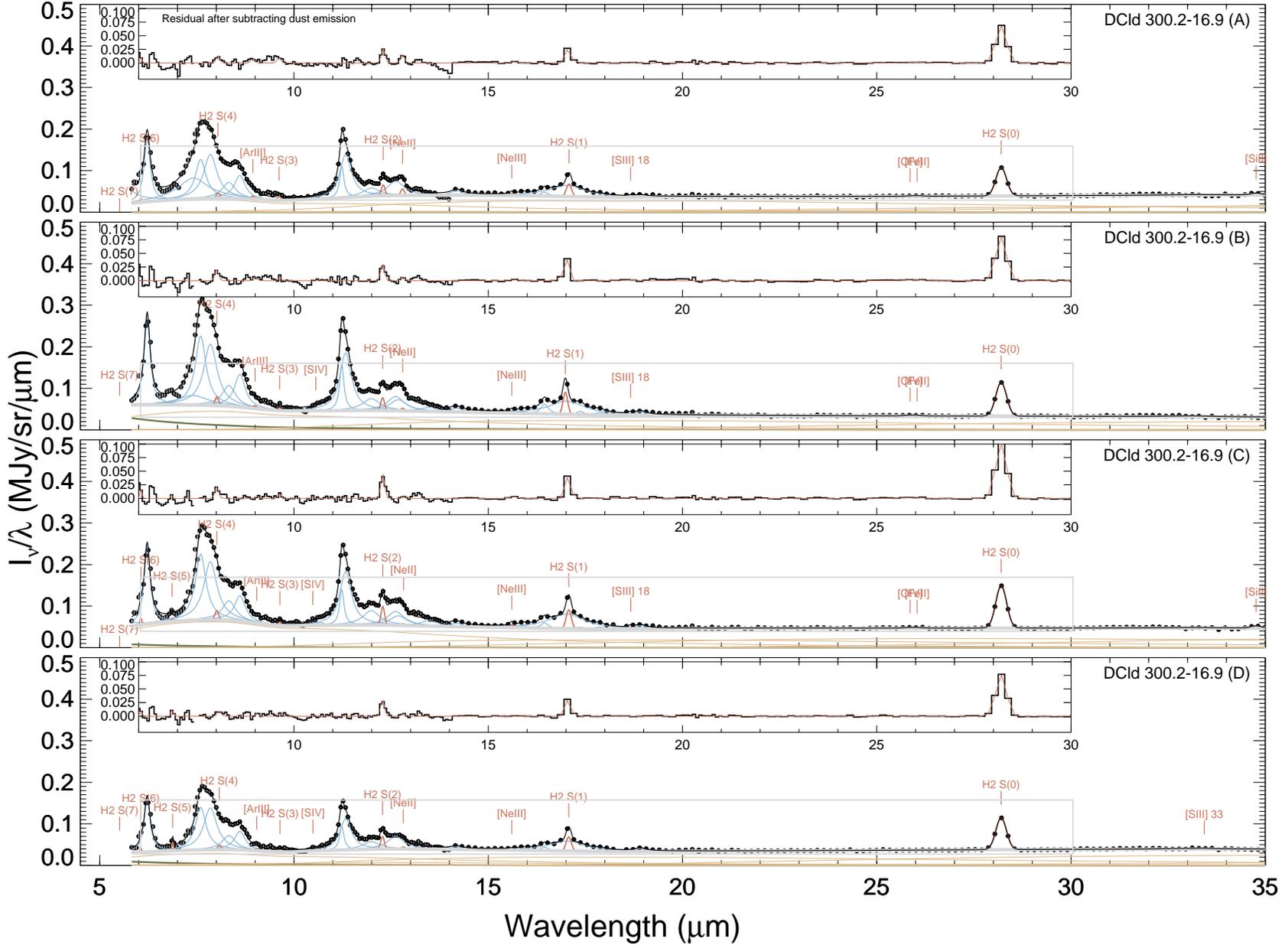}
\caption{\Spitzer\ IRS observed spectra (filled circles with error bars) for 4 positions in DCld 300.2-16.9 in front of the Chamaeleon complex.  Superimposed on the spectra are fits (solid black line) made using the PAHFIT program \citep{smi07}, that are each the sum of three dust continuum components (brown), a 5000\,K stellar blackbody (green), gas phase spectral lines (red), and a set of PAH features (blue), and account for silicate extinction. The thick gray curves indicate the total continuum emission.  The inset figures show the residuals in the data for a portion of the spectrum (solid histogram), after subtracting off the fits to the dust continuum and PAH features and correcting for extinction.  The summed fits to all gas phase spectral lines are superimposed in red on the inset plots.  All inset plot windows cover the same range in wavelength as the region of the main plot window they occupy.  The size of each inset window in its main plot window is given by a gray rectangle overlaid on the full spectrum. 
\label{cham_spectra}}
\end{sidewaysfigure*}

Our clouds all lie in regions of relatively bright Zodiacal emission [according to the {\sl Spitzer} Planning Observations Tool (SPOT)\footnote{http://ssc.spitzer.caltech.edu/warmmission/propkit/spot/} background estimator, the Zodiacal foreground at 7.7\micron\ ranges from 5--15\mjysr ]. To measure this sky foreground as well as correct for short term detector drifts, every observation (or concurrent group of observations) of a given cloud was accompanied by an off-source ``background'' measurement with the same integration parameters.  The background position was taken to be a local minimum in a $5\arcdeg\times 5\arcdeg$ {\sl IRAS} Sky Survey Atlas (ISSA) 100\micron\ map centered on the cloud.  The following sky background positions were observed (equatorial J2000): DCld 300.2--16.9 (11:41:33.7,--78:11:37); LDN 1780 (15:33:50.9,--07:28:26); MBM 12 (02:51:36.0,+19:10:03); MBM 28 (09:38:34.9,+71:13:58); LDN 183 (15:48:48.8, --00:34:35); and Stark 4 (23:52:41.1,--71:12:45).   
\vspace{1cm}
\subsection{Data Reduction}

Data processing started with the Basic Calibrated Data ({\tt bcd.fits}) spectral image files from the S18.7 \Spitzer\ IRS calibration pipeline.  As mentioned above, to maximize the signal to noise ratio in the SL data, we combined all 32 spectral images towards a given position, covering a $234\arcsec\times 3.7\arcsec$ region on the sky.  We used an outlier-resistant computation of the mean ({\tt resistant\_mean.pro} from the IDL Astronomy User's Library\footnote{http://idlastro.gsfc.nasa.gov/}) of the 32 measurements for each pixel.  For the LL data, we combined similarly the 48 spectral images covering the inner $280\arcsec\times 10.6\arcsec$ region covered by both subslits.  To remove Zodiacal foreground emission and residual dark current, and to minimize the effects of ``rogue'' pixels that vary on timescales greater than a few hours, we subtracted a mean image of the nearby sky (see above) from each mean source image.  We used IRSCLEAN\footnote{http://ssc.spitzer.caltech.edu/dataanalysistools/tools/irsclean/} to interpolate over any remaining rogue pixels in each of the sky-subtracted mean images.  We extracted full slit extended source calibrated spectra from the resulting images using the \Spitzer\ IRS Custom Extraction tool (SPICE).\footnote{http://ssc.spitzer.caltech.edu/dataanalysistools/tools/spice/}    

We combined the surface brightness spectra for SL orders 1 and 2 into a single spectrum covering the wavelength range 5.79--14\micron.  To do this we re-gridded the $\Delta\lambda = 0.04\micron$ SL2 spectrum to match the $0.06\micron$ SL1 wavelength grid using linear interpolation, and averaged SL1 and SL2 values where the two orders covered the same wavelengths.  Before combining, the edge of each order was trimmed by from 2--15 spectral elements to eliminate regions of poor responsivity.  This process was repeated for the LL data, giving a combined LL spectrum with a resolution of 0.17\micron\ covering the wavelength range 14.2-35.8\micron.  To obtain a 5.79--35.8\micron\ spectrum for the four positions with LL measurements, the SL and LL spectra for these four positions were concatenated together, retaining the different resolutions.   We used the PAHFIT package \citep{smi07} to fit simultaneously a set of Polycyclic Aromatic Hydrocarbon (PAH) dust features, gas phase lines (including the \htwo\ pure-rotational lines), dust continuum emission, and an extinction curve.  The fits were weighted by the error spectrum produced by the \Spitzer\ IRS pipeline and SPICE.  Uncertainties for all fit parameters were computed by PAHFIT using full covariance matrices for the multicomponent fits.  A proper fit to the data required modification of the default PAHFIT parameters.  The 12.69\micron\ PAH feature seems to be considerably wider in these spectra than expected from PAHFIT default parameters, resulting in spurious detections of the [Ne~{\sc ii}] 12.8\micron\ line when the default narrow 12.69\micron\ width was used.  This left a ``knee'' of emission in the model spectrum that was not visible in the data.  We found that increasing the fractional FWHM of the 12.69\micron\ Drude profile, from $\gamma_r (12.69) = 0.013$ to 0.042 (the same as for the 12.62\micron\ feature), removed the spurious knee and eliminated the [Ne~{\sc ii}] ``detection.''

For the S(3) through S(6) \htwo\ lines, which were not detected, we estimated 3$\sigma$ upper limits to the integrated intensity as follows.  First, we subtracted the fit to all PAH features and dust continuum components from the observed spectrum.  We divided the result by the PAHFIT-derived extinction curve (which in most cases did not affect the S(2) or 7.9\micron\ intensities by more than 10\%).  For a given \htwo\ line we then determined $\sigma_{\rmsmall rms}$, the root mean square of this residual extinction-corrected spectrum in a wavelength region centered on the line center wavelength, $\lambda_0\pm 4\times {\rm FWHM}_\lambda$.  Since all molecular lines are unresolved by the IRS, the expected full width at half maximum (FWHM$_\lambda$) of the line is determined only by instrumental parameters \citep[see][for more details on the \htwo\ line parameters]{smi07}.  For a gaussian line profile with peak intensity $I_{\lambda,{\rmsmall peak}}$ the integrated intensity is given by
\begin{align}
I \equiv \int{I_\lambda d\lambda} &= \frac{\sqrt{2\pi} ({\rm FWHM}_\lambda)}{2\sqrt{2 \ln 2}} I_{\lambda,{\rmsmall peak}} \nonumber
\\ &\approx 1.064 \,({\rm FWHM}_\lambda)\, I_{\lambda,{\rmsmall peak}}.
\label{ilambda_peak}
\end{align}
Thus the 3$\sigma$ integrated surface brightness limit is determined by substituting 3$\,\sigma_{\rmsmall rms}$ for $I_{\lambda,{\rmsmall peak}}$ in Equation \ref{ilambda_peak}:
\begin{align}
\frac{I}{\wmsr} &< 3.191\times\expo{-6}\,\left(\frac{\lambda_0}{\micron}\right)^{-2}\,\left(\frac{{\rm FWHM}_\lambda}{\micron}\right)\nonumber
\\ &\times\left(\frac{3\,\sigma_{\rmsmall rms}}{\mjysr}\right).
\label{upper_limit}
\end{align}

\subsection{Dust Emission}
\begin{figure*}[t!]
\includegraphics[keepaspectratio=true,scale=0.75]{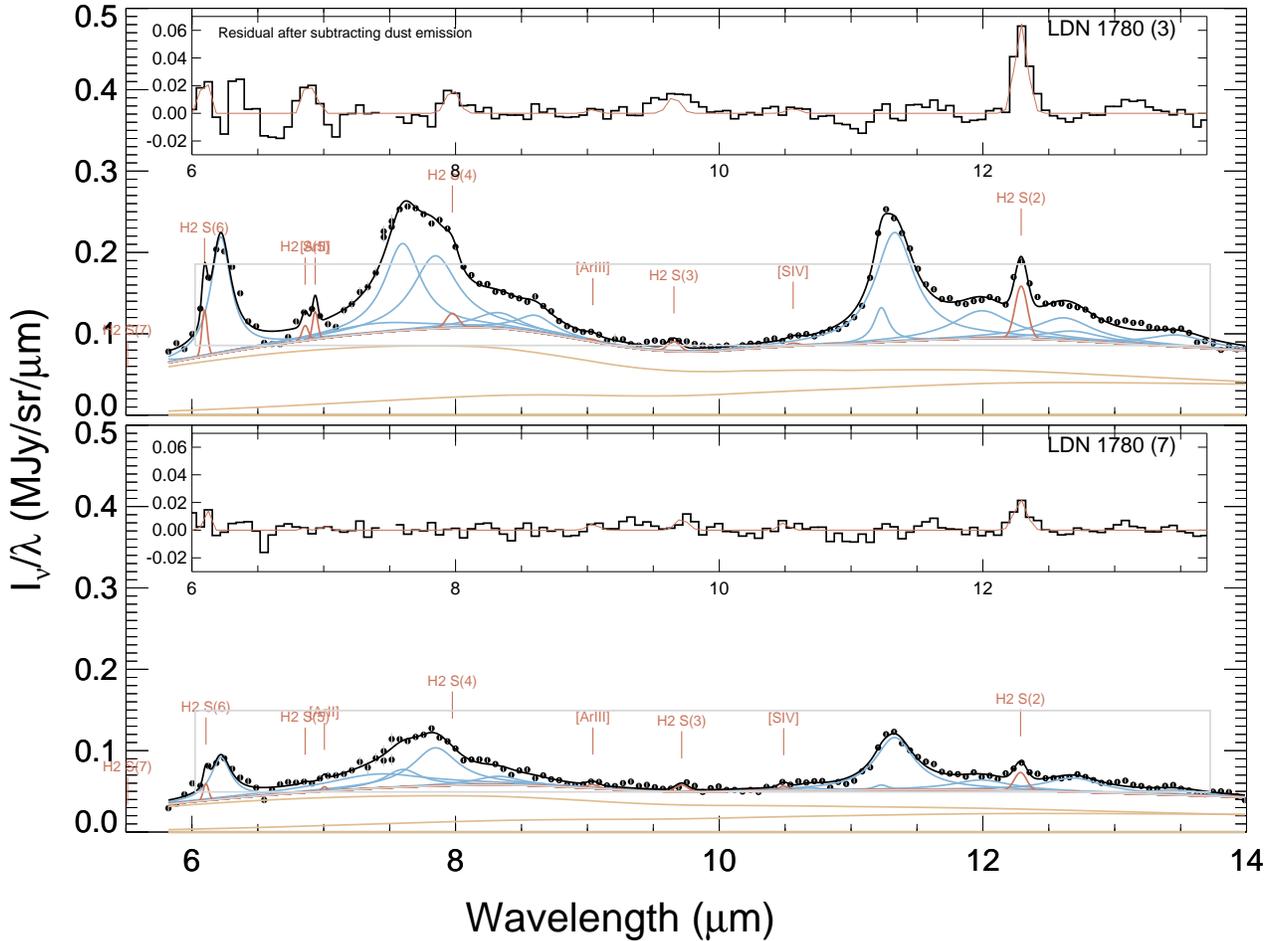}
\caption{Same as Figure \ref{cham_spectra}, except that these spectra are for 2 positions in LDN 1780, and only cover the SL wavelength range 5.8-14.0\micron.  Inset boxes cover a different range in wavelength and surface brightness per wavelength than in Figure \ref{cham_spectra} (but cover the same ranges in these two panels).
\label{mbm33_spectra}}
\end{figure*}

To constrain the amount of interstellar radiation processed by cirrus clouds, we synthesized broadband measurements of the dust emission.  For all positions, we computed $I_{\litl 7.9,PAH}=\langle\nu I_{\nu}\rangle_{\litl 7.9,PAH}$, the PAH surface brightness in \Spitzer\ Infrared Array Camera (IRAC) band 4, which is centered on 7.9\micron.  To do this we integrated the model PAH spectrum (given by the sum of fits to PAH features in Figures \ref{cham_spectra} and \ref{mbm33_spectra}) weighted by the IRAC 4 filter response, and multiplied the result by the color correction \citep{ssc10a} to give the specific intensity (\mjysr) that would be measured photometrically on an IRAC image.  Multiplying by the effective frequency of the 7.9\micron\ array gives the surface brightness (\wmsr).  This procedure was implemented using the \Spitzer\ Synthetic Photometry software.\footnote{http://ssc.spitzer.caltech.edu/dataanalysistools/cookbook/10/} Error bars for $I_{\litl 7.9,PAH}$ are derived from the covariance matrix for the fit PAH parameters.  For the position in Table \ref{nondetec_table} that did not have detectable PAH features we computed upper limits as 3 times the RMS spectrum integrated over IRAC band 4 (the spectrum was squared, integrated over the band, and multiplied by the color correction; the square root of this was multiplied by the effective frequency).  

We made a crude estimate of the total infrared (TIR) surface brightness to compare with the total \htwo\ emission from the S(0) to S(2) transitions.  We converted the formula given in Equation (22) of \citet{dra07a} from spectral luminosity to surface brightness: 
\begin{align}
{\rm TIR}&\approx 0.95\,\langle\nu I_{\nu}\rangle_{7.9} +
                                            1.15\,\langle\nu I_{\nu}\rangle_{24} \nonumber
                                            \\  & + \langle\nu I_{\nu}\rangle_{71}
                                            + \langle\nu I_{\nu}\rangle_{160} .
\label{tir_eqn}
\end{align}
The $\nu$ values are the effective frequencies (in Hz) of the \Spitzer\ IRAC 7.9 and MIPS 24, 71, and 160\micron\ bandpasses, and $I_{\nu}$ is the specific intensity measured with these instruments.  

The surface brightness $\langle\nu I_{\nu}\rangle_{7.9}$ was obtained as above.  For the four positions with LL spectra in DCld 300.2--16.9, $\langle\nu I_{\nu}\rangle_{24}$ was derived similarly using the LL spectrum and the MIPS 24\micron\ filter response, times an appropriate color correction \citep{ssc10b}. For the LDN 1780 positions, which did not have LL spectra, we estimated the 24\micron\ contribution to TIR as follows.  Taking the measured average value of $\langle \nu I_\nu\rangle_{7.9}$/TIR $=0.24\pm0.03$ for our sample, we used Figure 15 of \citet{dra07a} (second panel) to infer a dust mass fraction in PAHs of $q_{\litl PAH} = 4.6\%$.   For this value of $q_{\litl PAH}$, the third panel of the Figure predicts $\langle \nu I_\nu\rangle_{24}$/TIR $\approx 0.06$.  This assumes band emission relative to TIR does not vary as a function of illuminating radiation field strength $U$ (\ie\ $U\lesssim 10$, which probably holds for all cirrus clouds).  

Lacking 71 and 160\micron\ observations towards the DCld 300.2--16.9 positions, we used IRAS Sky Survey Atlas (ISSA) 60 and 100\micron\ measurements, respectively, as proxies for the missing MIPS photometry.  One of the HLCs in the full six-cloud sample, MBM-12 (LDN 1457)  (undetected in \htwo\ S(2) emission and not observed with LL), was mapped with MIPS by \citet{mag06}.  We computed the following rough scaling relations:
\begin{eqnarray}
I_{\nu,71} & = &  (2.30\pm 0.08)\,I_{\nu,60}\\
I_{\nu,160} & = & (4.98 \pm 0.13)\,I_{\nu,100}.
\end{eqnarray}
These were derived from robust linear fits to scatter plots of $I_{\nu,71}$ vs. $I_{\nu,60}$ and $I_{\nu,160}$ vs. $I_{\nu,100}$ for MBM-12, using images in the given wavebands resampled to the same pixel grids.   The errors in TIR are dominated by the 10\% accuracy of Equation \ref{tir_eqn}, as determined from model clouds heated by starlight with intensities between 0.1 and 100 times the average interstellar value \citep{dra07a}.   

\begin{figure*}
\plotone{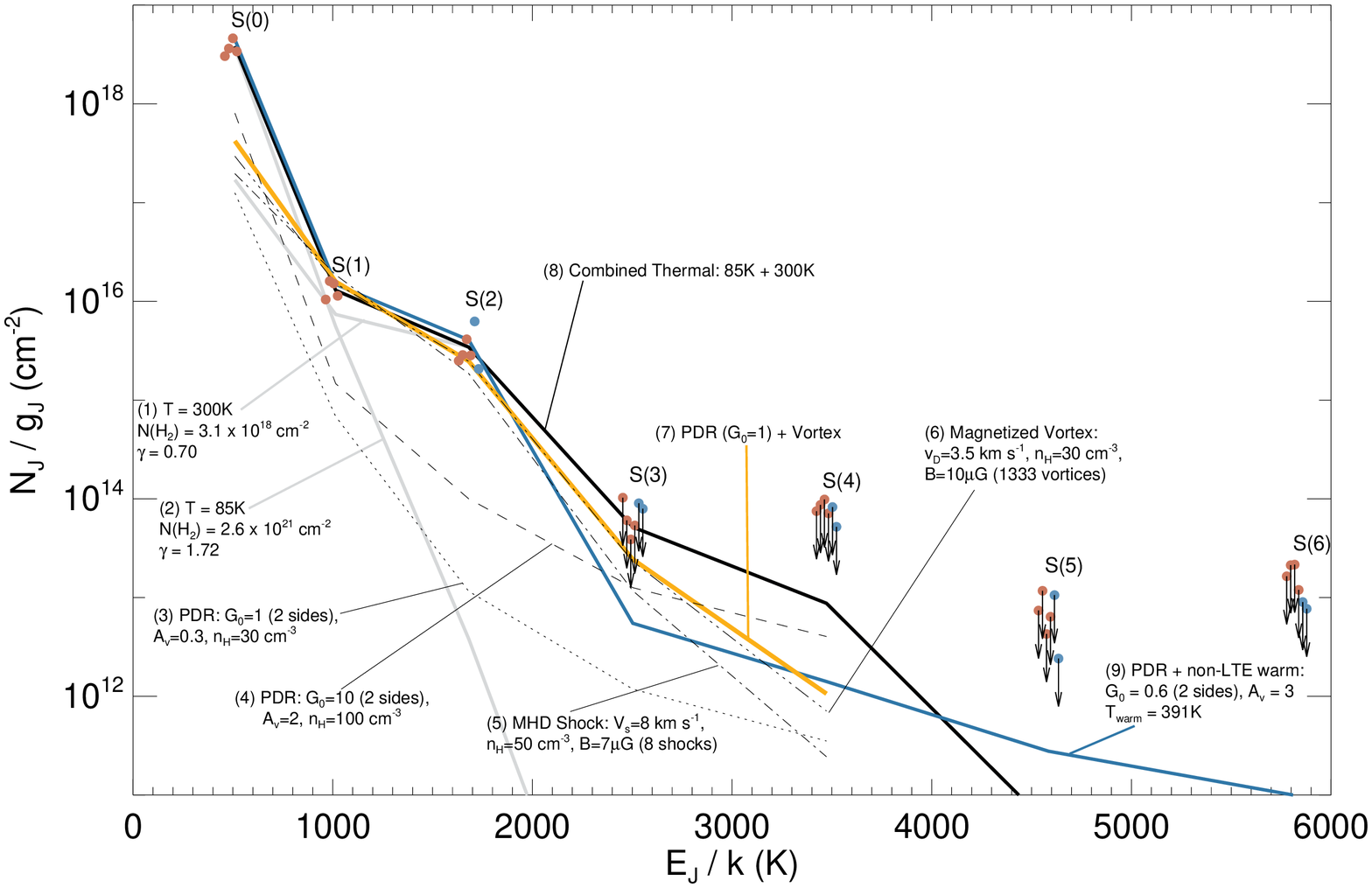}
\caption{Excitation diagram of \htwo\ pure-rotational emission towards high--latitude cirrus clouds.  Circles represent the observed measurements of DCld 300.2--16.9 ({\it red}) and LDN 1780 ({\it blue}).  Transitions S(3) and higher all have 3$\sigma$ upper limits, shown as downward arrows.  The abscissae for each point have been shifted slightly for clarity.  Superposed on the diagram are seven cloud models:  (1) a thermal distribution ($T_{\litl warm}=300\K$) with parametrized ortho/para ratio $\gamma=0.70$ ({\it gray}); (2) a thermal distribution ($T_{\litl cold}=85\K$) with $\gamma=1.72$ ({\it gray}); (3) emission from a PDR illuminated on both sides by a UV flux equal to the average local interstellar value \citep[$G_0=1$;][]{hab68}, volume density $\nh=30\cmthree$, and visual extinction $\av=0.3\,$mag ({\it dotted}); (4) PDR emission with $G_0=10$, $\nh=100\cmthree$, and $\av=2.0\,$mag ({\it dashed}); (5) emission from 8 MHD shocks traveling at $V_s=8\kms$ in gas with $\nh=50\cmthree$ and magnetic field $B=7\,\mu$G ({\it dot-dashed}); (6) emission from 1333 magnetized vortices with ion-neutral drift velocity $v_D = 3.5\kms$, $\nh=30\cmthree$, and $B=10\,\mu$G ({\it dot-dot-dashed}); (7) the sum of the $G_0=1$ PDR and the vortex model ({\it orange}); (8) the sum of the 85\K\ and 300\K\ thermal distributions ({\it solid black}); and (9) the sum of a PDR with $G_0=0.6$ plus a warm component with $T=391\K$, with non-LTE excitation.  Models (3)--(6) are taken from Fig. 3 of \citet{fal05}.  Model (9) is taken from Draine \& Ingalls (in preparation).
\label{hlc_h2_excitation}}
\end{figure*}

\section{Results}
\subsection{Measured Surface Brightnesses}
The complete 5.8--35.8\micron\ SL + LL spectra for the DCld 300.2--16.9 positions are displayed in Figure \ref{cham_spectra}.  The 5.8--14.0\micron\ SL spectra for the LDN 1780 positions are shown in Figure \ref{mbm33_spectra}.  We overlay on the spectra the fits to the grain and gas phase emission, and label the wavelengths of all lines in the model (regardless of detection).  We detected emission in the ($v$=0--0) 28.2\micron\ S(0) and 17.0\micron\ S(1) transitions of \htwo\ towards all four DCld 300.2--16.9 positions observed with the \Spitzer\ IRS Long Low (LL) spectrograph.  We detected emission in the 12.3\micron\ S(2) line at the $>3\sigma$ level in only six of the 34 HLC positions observed with the Short Low (SL) spectrograph:  the four in DCld 300.2--16.9 observed with LL and two in LDN 1780.  The 3$\sigma$ detection threshold for S(2) was between 0.7 and $5.4\times\expo{-10}\wmsr$ for the 34 observations.  For the cloud positions with detectable S(2) emission, the S(0), S(1), and S(2) surface brightnesses and their uncertainties (computed by PAHFIT) are listed in Table \ref{srctable}, columns 4 through 6.   

The S(3), S(4), S(5), and S(6) lines of \htwo\ were all too faint to be detected:  the integrated line strengths derived from attempting to fit the lines at the expected wavelengths were smaller than the corresponding errors in the fits.  Upper limits (3$\sigma$) to the surface brightness in these transitions (Equation \ref{upper_limit}) average about 4.2, 5.0, 5.1, and 6.8 $\times\expo{-10}\wmsr$, respectively.  We print the actual limits in columns 7--10 of Table \ref{srctable}.  For the positions without detectable \htwo\ emission, upper limits to the S(2) through S(6) intensities are listed in columns 4--8 of Table \ref{nondetec_table}.  

We list in Table \ref{srctable} the dust emission surface brightness: $I_{\litl 7.9,PAH}$ is in column 11, and the TIR estimates are in column 12.  For the positions without detectable \htwo\ emission, column 9 of Table \ref{nondetec_table} provides $I_{\litl 7.9,PAH}$ measurements.
\begin{figure*}
\plotone{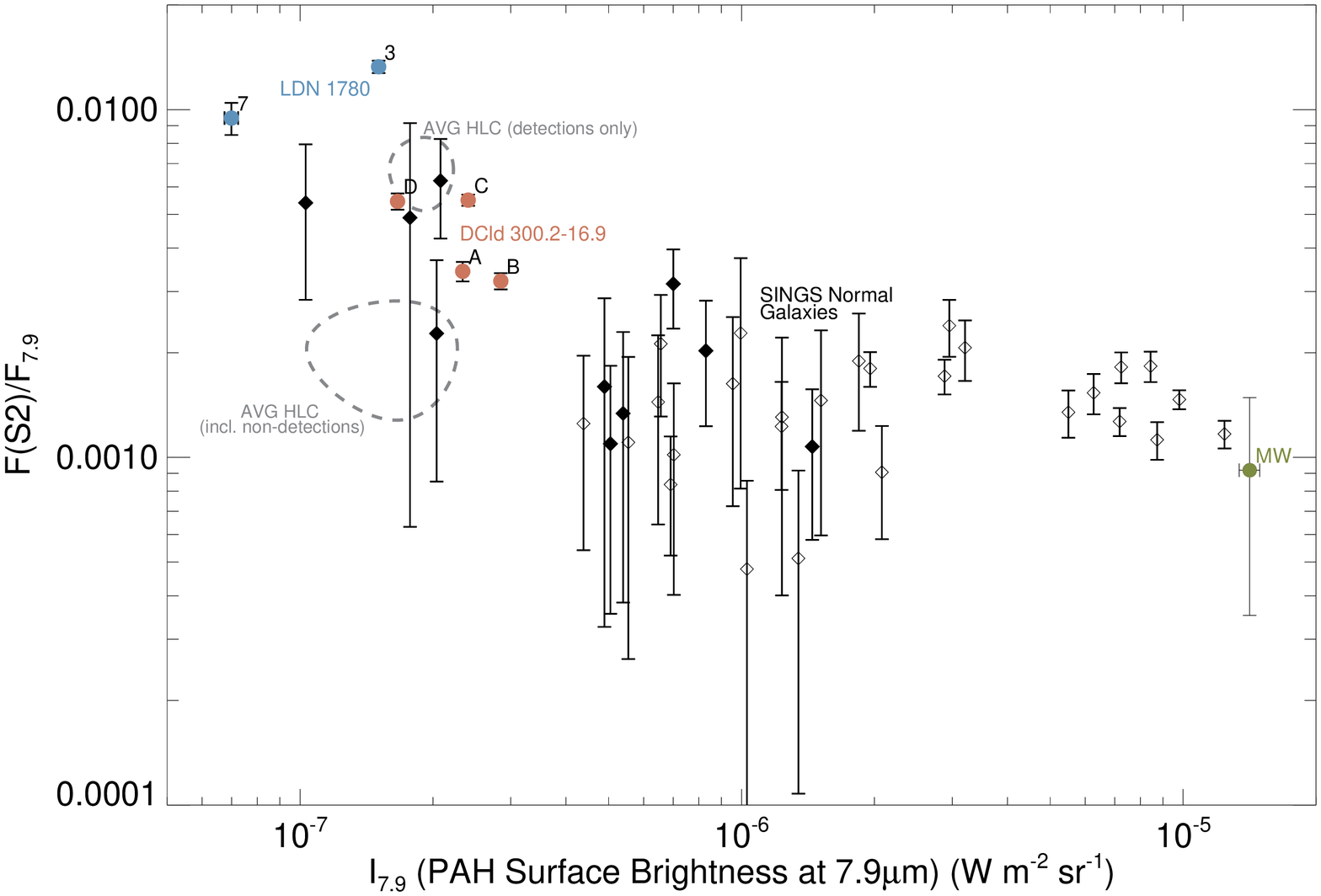}
\caption{Ratio of the \htwo\ S(2) to 7.9\micron\ PAH flux as a function of the 7.9\micron\ surface brightness, for Milky Way and extragalactic sources.  Red circles ({\it A-D}) are our DCld 300.2--16.9 HLC cirrus detections; blue circles ({\sl 3,7}) are the LDN 1780 detections.  The diamonds are \citet{rou07} data for SINGS galaxies with non-active nuclei (filled diamonds are dwarf galaxies and open diamonds are galaxies with purely star-forming nuclei).  The green circle (MW) is an estimate for the \citet{fal05} Milky Way galactic plane line of sight ($\av=30$).  Dashed ellipses (pinched due to the logarithmic scaling of the axes) indicate the average $\pm 1\sigma$ for two samples of high--latitude clouds: (1) the 6 S(2)--detected positions; and (2) the full set of 34 positions observed with SL, including non-detections.  
\label{hlc_roussel_h2}}

\end{figure*}
\begin{figure*}
\plotone{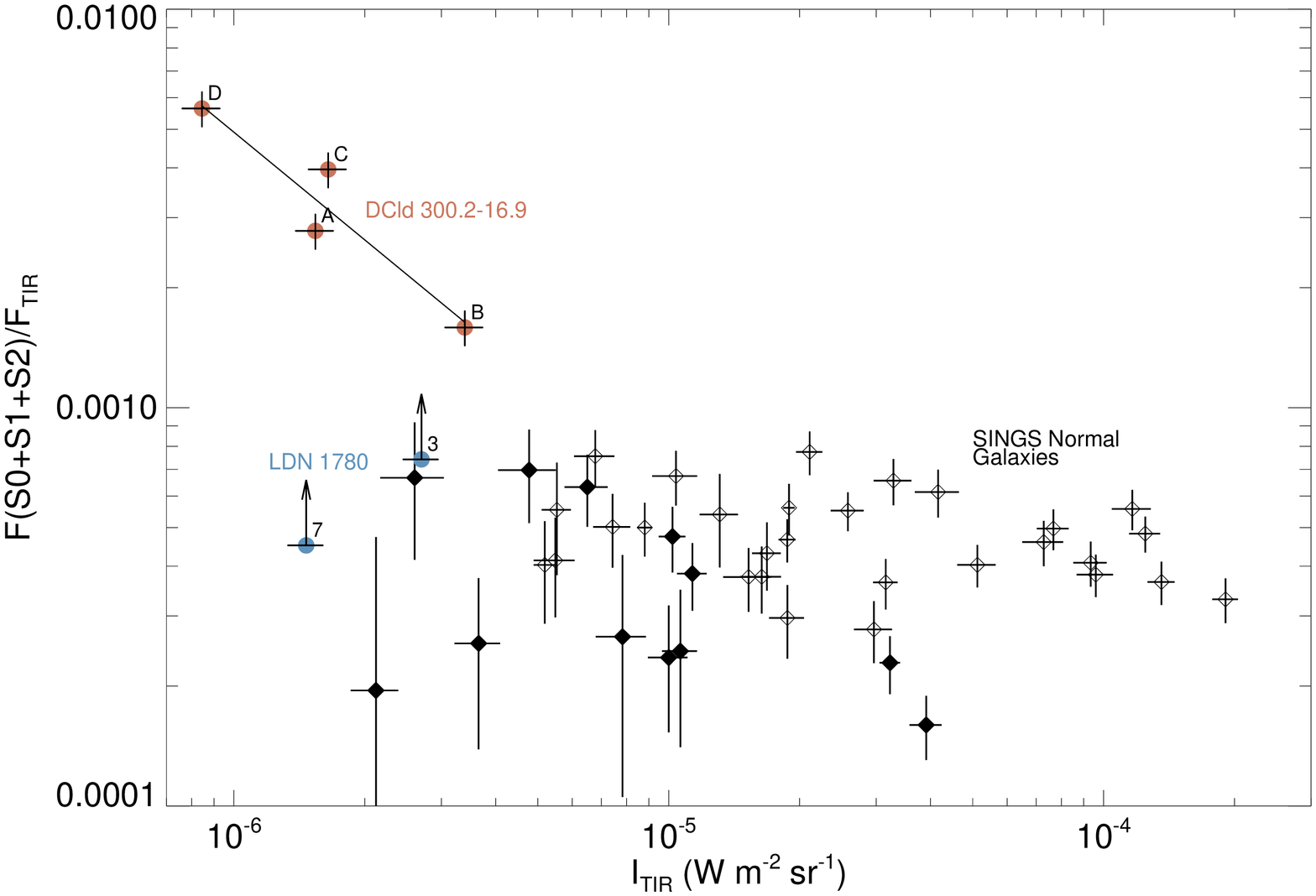}
\caption{Ratio of the total rotational \htwo\ flux to TIR flux, as a function of the TIR surface brightness.  Red circles ({\it A-D}) are our DCld 300.2--16.9 cirrus detections, blue circles (3 and 7) are lower limits for LDN 1780, and diamonds are \citet{rou07} data for SINGS galaxies with non-active nuclei (see Fig. \ref{hlc_roussel_h2}).  A power law fit to the DCld 300.2--16.9 points (gray line) yields $F({\rm S0 + S1 + S2})/F_{\litl TIR} = 2.0\times\expo{-8}\,{\rm TIR}^{-0.9}$.  
\label{hlc_roussel_tir}}
\end{figure*}

\subsection{Molecular Hydrogen Excitation}
We can use an excitation diagram to estimate the physical conditions in the regions producing \htwo\ rotational line emission in high--latitude translucent cirrus clouds.  In Figure \ref{hlc_h2_excitation}, we plot for all the Table \ref{srctable} measurements the column density in rotational upper level $J$ divided by the statistical weight, $N_J/g_J$, as a function of $E_J/k$, the upper level energy expressed as a temperature.  The column densities are derived from the surface brightness assuming optically thin line radiation:
\begin{equation}
N_J = \frac{4\pi\,I_{J,J-2}}{A_{J,J-2}\,h\nu_{J,J-2}}.
\label{col_dens}
\end{equation}
Here, $I_{J,J-2}$ is the surface brightness and $A_{J,J-2}$ is the Einstein spontaneous emission coefficient in the transition from upper state $J$ to lower state $J-2$.  The statistical weights for rotational level $J$ are $g_{J{\litl,even}} = 2J + 1$ for even $J$ (para-\htwo) and $g_{J{\litl,odd}}=3(2J+1)$ for odd $J$ (ortho-\htwo).  In what follows we assume that excitation of vibrational levels is negligible.  

Following \citet{dra96}, we approximate the \htwo\ rotational populations by a thermal distribution,
\begin{equation}
N_J/g_J = a\,\exp(-E_J/k\trot),
\label{boltzmann}
\end{equation}
where \trot\ is the ``rotational temperature'' describing the populations.  Under conditions of local thermodynamic equilibrium (LTE) Equation \ref{boltzmann} becomes a Boltzmann distribution, and $a = N(\htwo)/Z(\trot)$, the ratio of total \htwo\ column density to the partition function.  Under non-equilibrium conditions, the ortho- and para-\htwo\ can take on separate $a$ values, $a_{\litl ortho}$ and $a_{\litl para}$.  One way to characterize the relative populations in ortho and para states is the \citet{dra96} ortho to para ratio
\begin{equation}
\gamma = 3a_{\litl ortho}/a_{\litl para}.
\label{ortho_para}
\end{equation}
As those authors note, $\gamma$ is not necessarily the column density ratio of ortho- and para-\htwo, but rather a parametrization of that ratio that includes the ratio of partition functions of the two types of molecular hydrogen.  It has the advantage that under conditions of LTE ($a_{\litl ortho} = a_{\litl para})$, $\gamma = 3$, regardless of temperature.  

The observed excitation diagram can be used to estimate physical conditions in the S(0) to S(2)-emitting gas.  Using Equation \ref{boltzmann}, we note that for constant values of \trot\ and $a$, a plot of $N_J/g_J$ versus $E_J/k$ with logarithmically scaled ordinate will yield a straight line with slope equal to $-1/\trot$ (multiplied by $1/\log_{10}(e)$ if the ordinate is log base 10).  As is typical for other measurements of \htwo\ rotational populations in diffuse gas \citep[eg.,][]{rac01,fal05,gol10,jen10}, our data do not fall on a single straight line.  Here we simply determine two excitation temperatures consistent with the three lines available.  The result is somewhat arbitrary and not unique, but nevertheless is instructive.  

We find that a ``cold'' component with excitation temperature $T_{\litl cold} = 85\K$ and a ``warm'' component with $T_{\litl warm} = 300\K$ are consistent with the data.  This simple model, with $N(\htwo) = \nexpo{2.6}{21}\cmtwo$ in the cold component and $N(\htwo) = \nexpo{3.1}{18}\cmtwo$ in the warm component, reproduces the level populations in the $J = 2$, 3, and 4 levels (corresponding to the S(0), S(1), and S(2) transitions) and the upper limits in $J=5$ and above.  We overlay on Figure \ref{hlc_h2_excitation} the excitation diagram for each thermal component (gray curves), as well as the sum of the two components (solid black).  The model predicts a $J=5$ population that is right at the upper limit we have measured.  This is not a concern, because the critical densities for $J=3$, 4, and 5 are high enough that the $J=5$ level will be sub-thermally excited (at $T=70\K$, $n_{\litl crit}(\htwo) = 900$, $\nexpo{2}{4}$, and $\nexpo{3}{5}\cmthree$ for $J=3$, 4, and 5).  Thus the gas temperature would actually have to be higher than $T=300\K$ to reproduce the populations of $J=5$ predicted by our thermal model.  

In the model, we assume an ortho to para column density ratio of $N_{\litl ortho}/N_{\litl para} = 0.7$ for both the cold and warm components.  This is based on the observed column density ratio between the $J=1$ and 0 levels towards HD 102065 \citep{neh08b}, about 3\pdeg4 away from the nearest position in Table \ref{srctable}.  It seems reasonable to take the same ortho/para column density ratio for both components, on the theory that it is the cool gas that occasionally gets heated to $T = 300\K$, keeping the relative amounts of ortho and para \htwo\ constant. The {\it parametrized} ratio, $\gamma$, which measures departures from equilibrium (defined by $\gamma = 3$), is obtained by summing the predicted odd and even $J$ Boltzmann factors over all possible rotational levels (including $J=0$ and 1, which do not radiate), and taking the ratio.   This yields $\gamma_{\litl cold} = 1.72$ and $\gamma_{\litl warm} = 0.70$ for our thermal model.  

The two-temperature model ($T_{\litl cold}=85\K$ and $T_{\litl warm}=300\K$) presented here is an over-simplification.  In reality there will be a range of temperatures present, with much of the \htwo\ column density at $T< 85\K$ \citep[based on {\sl FUSE} absorption measurements towards 38 translucent lines of sight, the mean $J=1$ to $J=0$ rotational temperature is 67\K; see][]{rac09}, and some material at temperatures between 85\K\ and $T_{\litl warm}$ required to account for the observed excitation of $J=2$.

\subsection{Comparison with Measurements of the Disks of Galaxies}
To answer the question of whether translucent gas is a significant source of galaxian \htwo\ emission, we compare the \htwo\ and dust emission measurements with the \citet{rou07} SINGS galaxy sample and a measurement towards the mid-plane of the Milky Way.  We plot the flux ratios F(S2)/$F_{\litl 7.9,PAH}$ as a function of the surface brightness $I_{\litl 7.9,PAH}$ for the 6 HLC positions in Figure \ref{hlc_roussel_h2}.  (Since the solid angles of the \htwo\ and PAH measurements are the same, the surface brightness ratio equals the flux ratio.)  

We also plot the SINGS extragalactic measurements of F(S2)/$F_{\litl 7.9,PAH}$ vs. $I_{\litl 7.9,PAH}$ in galaxies without active nuclei \citep{rou07}.  We derived 7.9\micron\ surface brightnesses for this sample by dividing the reported \citeauthor{rou07} IRAC band 4 fluxes corrected for stellar emission by the solid angle for each galaxy observation.  

We add a final data point to Figure \ref{hlc_roussel_h2}:  the ISO SWS detection of S(2) along a direction in the mid-plane of the Milky Way \citep[MW; $\ell,b$=26\pdeg46,0\pdeg09;][]{fal05}.  Lacking an IRAC 7.9\micron\ measurement or its equivalent for this position, we scaled the IRAS 12\micron\ measurements \citep[IRIS processing;][]{miv05} for the MW position to the average $7.9\micron/12\micron$ ratio derived from HLC synthetic photometry (7.9\micron) and IRIS measurements (12\micron).  We fit a least squares bisector line \citep{iso90} to the 7.9 and 12\micron\ measurements and used the slope as the nominal scale factor, $I_{\litl 7.9,PAH}/I_{\litl 12,IRIS}$, to multiply by the 12\micron\ MW measurement.  We estimated the error in the predicted $I_{\litl 7.9,PAH}$(MW) value using the standard deviation in the six measured HLC values of the $I_{\litl 7.9,PAH}/I_{\litl 12,IRIS}$ ratio (after subtracting the bisector fit $y$-intercepts). Multiplying this by $I_{\litl 12,IRIS}$(MW) gives an error in $I_{\litl 7.9,PAH}$ (MW).  This error is probably an underestimate, since it ignores possible systematic differences in the PAH spectra of HLC and MW populations. 

The six HLC positions have higher values of the F(S2)/$F_{\litl 7.9,PAH}$ ratio than the Milky Way mid-plane and all but three of the SINGS non-active galaxies. The average F(S2)/$F_{\litl 7.9,PAH}$ ratio for the detected sample is $(6.7\pm 1.6)\times\expo{-3}$.  For the \citet{rou07} SINGS sample the ratio averages $(2.9\pm 0.6)\times\expo{-3}$ for dwarf galaxies, $(1.44\pm 0.09)\times\expo{-3}$ for galaxies with purely star-forming nuclei, and $(1.8\pm 0.2)\times\expo{-3}$ for the combined set.  For the \citet{fal05} Milky Way position, the ratio is $(1.0\pm 0.6)\times\expo{-3}$.  For the complete 34-position HLC sample, 27 lines of sight showed PAH emission in the 7.9\micron\ band but no S(2) emission (Table \ref{nondetec_table}).  If we include all 7.9\micron\ detections in the HLC average, the ensemble has a ratio of $(2.0\pm 0.8)\times\expo{-3}$, which is identical within the error bars to the full SINGS average and the Milky Way data point.  (We estimated the average S(2) intensity for the complete sample using a spectrum produced by coadding the background-subtracted IRS spectral images for all 34 positions.)  In the following section we discuss the possibility that translucent cirrus comprises a significant component of galaxian \htwo\ rotational emission.

A useful diagnostic of the relative strength of \htwo\ emission to absorbed interstellar radiation is the ratio of the flux in the S(0), S(1), and S(2) lines to the total infrared flux, $F({\rm S0+S1+S2})/F_{\litl TIR}$.  We plot in Figure \ref{hlc_roussel_tir} this ratio as a function of $I_{\litl TIR}$ for the four DCld 300.2--16.9 positions with LL spectra, as well as the two lower limits from the LDN 1780 S(2) data.  Here the DCld 300.2--16.9 points lie well above the extragalactic measurements.  The average value of $F({\rm S0+S1+S2})/F_{\litl TIR}$ for DCld 300.2--16.9 is $(3.5\pm 0.9)\times\expo{-3}$.  For the SINGS data, the ratio is $(3.7\pm 0.6)\times\expo{-4}$ for dwarf nuclei, $(4.8\pm 0.2)\times\expo{-4}$ for pure star-forming nuclei, and $(4.5\pm 0.8)\times\expo{-4}$ for the complete sample.  It is notable that for the LDN 1780 positions, $F({\rm S2})/F_{\litl TIR}$ alone is of the same order of magnitude as $F({\rm S0+S1+S2})/F_{\litl TIR}$ for the SINGS galaxies.  Unlike extragalactic \htwo\ emission, the \htwo\ emission for HLCs seems only loosely correlated (if at all) with the infrared flux:  a power law fit to the DCld 300.2--16.9 data in Figure \ref{hlc_roussel_tir} has $I(S0+S1+S2) \propto I_{\litl TIR}^{0.1}$. 

\section{Discussion}
We have detected molecular hydrogen rotational emission towards six of 34 translucent high Galactic latitude cirrus positions.  The ($v=0-0$) S(0), S(1) , and S(2) transitions of \htwo\ originate from energy levels that are 510\K, 1015\K, and 1682\K\ above the ground state.  How are these levels excited in the translucent ISM, which is embedded in cold \hi\ gas with a median temperature of $\sim 70\K$ \citep{hei03}?  Why do some locations show S(2) emission and others do not?  Do we expect a significant amount of \htwo\ cirrus emission throughout the Galaxy, and by  extension in other normal galaxies?  These are the main questions we will try to address here.

\subsection{The Environment of the Clouds}
We first examine the environment of the two clouds we have detected, DCld 300.2--16.9 and LDN 1780, focusing on aspects of the clouds that may explain the \htwo\ emission we have observed.  Both clouds are thought to be part of the cold neutral medium (CNM) that defines the boundary of the 30--100\pc\ diameter Local Cavity (LC) of low-density ($n\lesssim 0.1\cmthree$) gas in which the Sun resides \citep{wel10}.  The clouds also  both happen to be on the interface between the LC and the superbubble surrounding the Scorpius-Centaurus (Sco-Cen) OB association \citep{deg92}, albeit on opposite sides.  Southeast of the superbubble center (in Galactic coordinates), DCld 300.2--16.9 is located along the line of sight to the Chamaeleon molecular cloud, but is probably closer than the $\sim$150 pc distance of the main complex \citep{miz01,neh08a}.  \citet{miz01} deduced that the cloud is at a distance of $70\pm 15\pc$ from the Sun, based on the location of T Tauri star T Cha, towards which the cloud peak of CO emission has the same radial velocity.  The cloud is thus $\sim 30\pc$ from the center of the Lower Centaurus Crux (LCC) subgroup of Sco-Cen.  Northwest of the superbubble center, LDN 1780 ($\ell,\,b = 358\pdeg9,\,36\pdeg 9$) is about 100\pc\ from the Sun \citep{fra89,lal03} and is thus $\approx 70\pc$ away from the center of the Upper Scorpius (US) subgroup of Sco-Cen \citep{lau95, deg92}.  Arrows pointing from the clouds to the centers of the LCC and US subgroups are drawn on Figure \ref{slit_geom}.  Projected on the plane of the sky, LDN 1780 is $\approx 75\arcdeg$ away from DCld 300.2--16.9, so the clouds are separated by at least 100\pc.  

In terms of their processing of the interstellar radiation field (ISRF), both clouds have 60/100\micron\ colors and \cii\ cooling intensities that are typical of translucent Galactic cirrus, and have been modeled as objects with visual extinctions $\av\sim 1$ magnitude and ISRF intensities $G_0\sim 1$, in units of the \citet{hab68} flux \citep{ing02,juv03,neh08b}.  

We estimate that the Table \ref{srctable} positions have extinctions of from $1-3$\,mag.  For DCld 300.2--16.9, the \citet{miz01} \co\ map has integrated intensities $I_{\litl CO}\sim 10\K\kms$ near our positions, which yields a molecular column density of $N(\htwo)\sim 2\times\expo{21}\cmtwo$, if we use a typical conversion factor between $I_{\litl CO}$ and $N(\htwo)$ \citep[eg., see][]{lis10}.  Using the well-established relationships $\nnh/E_{B-V} = 5.8\times\expo{21}\cmtwo$ and $\av/E_{B-V} = 3.1$ in diffuse gas \citep{boh78,rac09}, this gives $\av \approx 1.1\,$mag (this is probably an underestimate since it does not count \hi).  The LDN 1780 positions have measured visual extinctions of $\av\approx 2-3\,$mag \citep{del06}, which corresponds to a total gas column density of $\nnh\sim (3.7-5.6)\times\expo{21}\cmtwo$.  Alternatively, the average 100\micron\ surface brightness towards all detected DCld 300.2--16.9 and LDN 1780 positions is 13.4\mjysr.  Using the \citet{bou96} conversion $N($\hi$) \approx \nexpo{2}{20}\cmtwo \, (I_{100}/\mjysr)$ yields a column density of $\nexpo{2.7}{21}\cmtwo$, or $\av = 1.4\,$mag.

The portion of DCld 300.2--16.9 in front of HD 102065, about $3\pdeg4$ (4.2\pc) away from the nearest position in Table \ref{srctable}, has been studied in absorption of \htwo, CH$^+$, and many other species by \citet{gry02} and \citet{neh08a}.  A model of the direction towards this star can reproduce most of the observations if $\av\sim 0.7$ and $G_0 \sim 0.7$  \citep{neh08b}.  However, the populations in the \htwo\ $J>2$ rotational levels require excitation beyond that produced by any reasonable incident radiation field to explain the observed absorption.  The authors concluded that a warm ($T\sim 250\K$), out-of-equilibrium gas is also present along the line of sight.  \citet{neh08a} posited that the cloud's current state is the result of an interaction between a supernova blast wave from one or more stars in the LCC subgroup and the ambient ISM, about $2-3\times\expo{5}\,$yr ago.   

LDN 1780 is also thought to have been affected profoundly by its proximity to the Sco-Cen association.  The cloud is located between two near-parallel arcs of \hi\ emission. The outer arc is assumed to be \hi\ swept up by supernova explosions in the Upper Centaurus Lupus (UCL) subgroup of Sco-Cen, whereas the inner arc is believed to have been produced by the supernovae in the Upper Scorpius (US) subgroup \citep{deg92}.  \citet{tot95} proposed a scenario in which a shock front propagating outward at a speed of $\sim 10-15\kms$ from UCL has passed through LDN 1780.  Since the loop is currently about 7\pc\ away from the cloud, we estimate that the cloud's interaction with the shock probably occurred $\sim 6\times\expo{5}\,$yr ago.  \citeauthor{tot95} noted that the \hi\ emission from LDN 1780 peaks on the lower latitude diffuse side of the cloud facing the US subgroup and the stars of the Galactic mid-plane; the CO emission peaks on the dense side of the cloud further from these sources of radiation. They argued that the cometary shape of the cloud is due to the passage of the UCL shock wave, whereas the segregation between molecular and atomic material is due to an asymmetric UV radiation field originating from the US subgroup and the Galactic mid-plane.  \citet{lau95} estimated that $G_0$ might be as high as 3 on the diffuse side of the cloud, but may be only half as strong on the dense side.  The \htwo\ S(2) emission we detect is stronger towards position 3 (Table \ref{srctable}), which is close to the \hi\ peak.  We argue in \S4.2 that the UV field is insufficient to produce the emission, and suggest instead that the S(2) emission results from the cloud's interaction with the supernova blast wave.

LDN 1780 is unusual among molecular clouds in that it has a significant H$\alpha$ surface brightness, correlated with tracers of column density \citep{del06,wit10}.  \citet{mat07} suggested that this could be explained by scattering of background H$\alpha$ photons by dust grains, and the recent study by \citet{wit10} strongly supports this interpretation.  The $\zeta$\,Oph and $\delta$\,Sco \hii\ regions appear to be within a few tens of pc of the southern surface of LDN 1780, and represent likely sources of background H$\alpha$.  

\subsection{Sources of Molecular Hydrogen Excitation}
In interstellar cirrus, \htwo\ may be excited into rotational emission via UV pumping by starlight; or via collisional excitation resulting from mechanical energy dissipation in magnetohydrodynamic (MHD) shocks \citep{dra83,flo86}, coherent vortices or intense velocity shears \citep{jou98,fal05,god09}, or by interaction with cosmic rays \citep{fer08}.  All of these sources may be present at some level in DCld 300.2--16.9 and LDN 1780.

\subsubsection{UV Pumping}
The interstellar radiation field (ISRF) is the most obvious source of \htwo\ excitation in cirrus clouds.  FUV radiation heats the gas and dust and drives the chemistry in all translucent regions.  \citet{rou07} found a correlation between \htwo\ emission and infrared emission in the disks of SINGS normal galaxies.  Since IR radiation traces the processing of starlight by dust, SINGS \htwo\ emission appears to be tied to the FUV radiation field.  But for cirrus clouds, it is difficult to reproduce the observed \htwo\ emission with only the ISRF.  We overlay on Fig. \ref{hlc_h2_excitation} the predicted excitation in two photodissociation region (PDR) models \citep{fal05}.  The first model is immersed in the \citet{hab68} interstellar radiation field ($G_0=1$) and has gas volume density $n = 30\cmthree$, and visual extinction $\av = 0.3\,$mag.  The second PDR model has a larger UV flux, and is denser and thicker: $G_0=10$, $n=100\cmthree$, and $\av=10\,$mag.  The translucent HLCs in our study are probably somewhere in between these models.  Both models fall short of reproducing our observed measurements, by at least a factor of 10. In another \htwo\ fluorescence model, that of \citet{bla87}, the S(2) emission we observe can only be reproduced if the radiation field was increased to $G_0\sim 30$.  This would, however, result in S(3) emission at a level we would have detected.  Furthermore, there is no evidence that these HLCs are heated by radiation fields greater than $G_0\approx 3$ \citep{lau95,ing02,juv03,neh08b}.  So another mechanism must dominate the \htwo\ pure-rotational excitation. 

We note that by far the brightest S(2) emission we have detected is towards position 3 of LDN 1780.  This position has 3 times as much S(2) radiation as position 7, and also has twice as much 7.9\micron\ PAH emission as position 7, consistent with the PAHs on the southern side of the cloud (in Galactic coordinates) being exposed to higher starlight intensities, as predicted by \citet{lau95}.  The ISRF would have to be increased by a factor of more than 100, however, to account for the \htwo\ S(2) emission towards this position, which would boost the PAH emission by a similar factor (provided the density remained the same) and would result in detectable S(3) emission and much brighter S(0) and S(1) lines than we observe.  The highest {\sl IRAS} 60/100\micron\ surface brightness ratio measured towards LDN 1780 is 0.275 \citep{ing02}, which implies $G_0\sim 2-4$.  \citet{wit10} found that for the entire cloud, grain temperatures are in the range 14.5--16.8\K, consistent with $G_0\sim1$.  Therefore it is likely that most of the S(2) radiation represents the dissipation of mechanical energy that was concentrated on the south side of the cloud (like the passage of a supernova blast wave).

\subsubsection{Collisional Excitation}
\htwo\ rotational transitions can be collisionally excited into emission via shock heating \citep{dra83} or turbulent dissipation \citep{god09,fal05}.  

What kind of shocks would be capable of powering the observed emission?  The observed \htwo\ S(0) to S(2) line intensities sum to $I(0-2) \equiv I[S(0)] + I[S(1)] + I[S(2)] \approx \nexpo{5}{-9}\wmsr$ (on average) at the 4 DCld 300.2--16.9 positions where emission is detected.  If a fraction $f(0-2)$ of the shock power is radiated in these lines then, assuming the shock front fills the beam, 
\begin{equation}
I(0-2) \approx \frac{1}{8\pi} \frac{1}{\cos\theta} f(0-2) \rho_0 v_s^3
\end{equation}
for a strong shock, where $\theta$ is the angle between the shock normal and the line of sight, $\rho_0$ is the density of the pre-shock gas, and $v_s$ is the shock propagation velocity.  Thus, 
\begin{align}
v_s &\approx 8\kms  \nonumber
\\ & \times \left[\frac{I(0-2)}{\nexpo{5}{-9}\wmsr}
\frac{\cos\theta}{f(0-2)}
\frac{100\cmthree}{{\nh}_{,0}}\right]^{1/3}.
\end{align}
Therefore, if the S(0-2) lines are major coolants, and if the line ratios are also consistent, a shock speed of order 8\kms\ would be sufficient to produce the observed line fluxes.  Figure 4 of \citet{dra83} indicates that the \htwo\ lines are in fact expected to dominate the cooling of $v_s=5-20\kms$ C-type shocks propagating into gas with ${\nh}_{,0}=100\cmthree$ and $B_0=10\,\mu$G.  

Matching the line ratios is a separate challenge.  The $J=3$ and $J=4$ levels could be excited in shocks.  \citet{dra86} show results for shocks propagating into gas with $\nh = 50\cmthree$ and $B_0 = 7\,\mu$G.  For $v_s \approx 12\kms$ and $\cos\theta = 0.5$, these models reproduce the observed S(2) emission and overestimate by a factor of about 2.6 the S(1) emission.  The models assumed an ortho/para ratio of 3; using a smaller ratio (eg., $\gamma = 0.7$) could decrease the S(1)/S(2) ratio to agree with the observed values.  

If such shocks are present in these clouds, one would expect to see 21\,cm emission with components corresponding to both the pre-shock and post-shock material.  CO ($J=1-0$) line emission should also show pre-shock and post-shock components.  Such signatures have yet to be detected.  

Published shock models fail to predict the S(0) emission we observe towards HLCs without overestimating the S(1) and/or S(2) lines.  The very large column density $N(J=2) = \nexpo{1.5}{19}\cmtwo$ appears to require a large filling factor of \htwo\ gas with relatively cool temperatures.  For thermal excitation we reproduce the S(0) emission if most of the gas has an excitation temperature of $T = 85\K$ (Fig. \ref{hlc_h2_excitation}, black curve). 

One approach to increasing the $J=2$ population relative to $J=3$ and 4, while still reproducing the overall column densities, has been to posit multiple superimposed, randomly oriented shock waves.  Both \citet{gre02} and \citet{fal05} applied the MHD shock model of \citet{flo98} to measurements of molecular lines in diffuse clouds and found that many of the observations could be explained by ensembles of shock waves.  For comparison with our data, we display on Fig. \ref{hlc_h2_excitation} (dash-dot curve) the sum of 8 shocks traveling with velocity $v_s = 8\kms$ into a medium of $\nh = 50\cmthree$ and $B=7\,\mu$G \citep[models taken from Figure 3 of][]{fal05}.  \citeauthor{fal05} found that, per magnitude of visual extinction, 3.2 shocks with $v_s=8\kms$ can reproduce their S(1) and higher emission from the diffuse medium.  This is nearly the same ``shock fraction'' we need for the S(1) and S(2) data towards DCld 300.2--16.9 and LDN 1780 (our sample has $\av\approx 2\,$mag, giving 4 shocks per magnitude).  Since the shocks each have a column density $N_{\litl sh}= \nexpo{1.5}{17}\cmtwo$ of hot gas, the mass fraction of shocked gas in this model is about 0.06\%.  The model $J=3$ and $J=4$ column densities are reasonably well-matched to the data, but $J=2$ is still insufficiently populated.  Indeed, referring to the \citet{dra86} models, even a $v_s=5\kms$ shock overproduces S(1) emission relative to S(0) by a factor of 8, so the $J=2$ levels remain a problem to reproduce for shock models.

Although shocks provide one mechanism for localized heating, it has also been proposed that intermittent dissipation of turbulence can produce localized, transient pockets of hot gas \citep{jou98,fal05,god09}.  We show on Fig. \ref{hlc_h2_excitation} the excitation diagram for 1300 vortices, scaled from Figure 3 of \citet{fal05} (dot-dot-dash curve).  The vortices have ion-neutral drift velocity $v_D = 3.5\kms$, $\nh = 30\cmthree$, and $B=10\,\mu$G.  The model has very similar \htwo\ excitation to the multiple shock model described above, although it has a lower ortho/para ratio which gives a slightly better match to our S(1) and S(2) data.  We get the same ``vortex fraction'' as \citeauthor{fal05}, $\sim 700$ vortices per magnitude of visual extinction.  The vortex column density of $N_{\litl vor} = \nexpo{1.7}{14}\cmtwo$ gives a mass fraction in the vortices of about 0.01\%.  

Since all cirrus is exposed to the ISRF, a more realistic model contains both UV and collisional heating.  We display on Fig. \ref{hlc_h2_excitation} the sum of the $G_0=1$ PDR and the $v_D=3.5\kms$ vortex models (orange curve).  The S(0) line remains underpredicted, by about a factor of 10.  

We have computed an additional model that reproduces the \htwo\ level populations quite well, even though it lacks a specific mechanism for heating the warm molecular gas (Draine \& Ingalls, in preparation).  This model is plotted in blue on Fig. \ref{hlc_h2_excitation}.  It consists of a plane-parallel molecular cloud with an \hi\ layer (PDR) on both front and back sides, superimposed with a zone of warm gas.  The PDR portion of the slab obtains a uniform gas pressure $P/k = 4000\cmthree\K$ (consistent with pressure measurements in the ISM based on [\ci ] excitation---see \citeauthor{jen11}2011), and is embedded in a radiation field equal to 0.6 times the \citet{hab68} energy density at 1000\AA.  The warm zone is assumed to have been suddenly heated, so that the ortho/para ratio has not had time to reach equilibrium corresponding to $T_{\litl warm}$.  Collisional excitation of the \htwo\ rotational levels is calculated using state-to-state rates.  The data are best fit by $T_{\litl warm} = 391\K$, an ortho/para column density ratio in both the warm and cool gas of 0.15, and a warm \htwo\ column density $N_{\litl warm}(\htwo)\approx \nexpo{5.15}{19}\cmtwo$, which is about 2\% of the total \htwo\ column density in the model.  The bulk of the \htwo\ emission in this model is not due to radiative pumping; more than 97\% of the $J=2$, 3, and 4 excitation is produced in the warm component.  

\subsection{\htwo\ Excitation in the Supernova-Driven ISM}
In summary, the (somewhat speculative and incomplete) scenario we envision is one in which CNM molecular clouds immersed in the interstellar radiation field of intensity $G_0\sim 1$ are subjected to a large scale injection of mechanical energy and dissipate that energy intermittently on smaller scales via shocks, or perhaps in localized transient turbulent dissipation regions \citep[TDRs;][]{god09} such as vortices or velocity shears (or some combination thereof).  The \htwo\ S(0) -- S(2) lines are presumed to be mainly excited by collisions in hot regions, heated by mechanical energy dissipation.  The gas is initially cold ($T\lesssim 100\K$), and in the shielded interior of a cloud the ortho/para ratio may be very low.  When dissipational heating occurs, local temperatures temporarily reach values near 1000\K, and ortho/para rises.  If the source of large scale mechanical energy is not refreshed somehow, the fraction of warm gas will decrease until the line radiation is no longer detectable.  

Since both clouds in this study are thought to have been affected by supernova explosions from the Sco-Cen OB association about $\nexpo{(2-6)}{5}\,$yr ago \citep[][]{tot95,neh08a}, we surmise that portions of the medium are still dissipating the energy deposited from those 100\pc\ scale blast waves.  For this scenario to hold, the time to dissipate the gas turbulent kinetic energy must be larger than the time since blast wave impact.  Using the total \htwo\ S($0-2$) intensity from the DCld 300.2--16.9 positions, $I(0-2)  \approx \nexpo{5}{-9}\wmsr$, we can calculate the \htwo\ cooling rate per H atom:
\begin{align}
\Lambda_{\rm H} & =  \frac{4\pi\,I(0-2)}{\nnh}\nonumber
    \\            & =  \nexpo{3.1}{-33}\,{\mathrm W}\,({\rm H~atom})^{-1} \nonumber
    \\            & \times \left(\frac{I(0-2)}{\nexpo{5}{-9}\wmsr}\right)\,\left(\frac{\nexpo{2}{21}\cmtwo}{\nnh}\right).
\end{align}
The turbulent kinetic energy per proton is
\begin{align}
E_{\litl kin,H} &= 1.4\times\frac{3}{2}\,m_p\,\sigma_v^2 \nonumber
\\              &= \nexpo{3.2}{-20}\,{\mathrm J}\,({\rm H~atom})^{-1}\,\left(\frac{\sigma_v}{3\kms}\right)^2,
\end{align}
where $m_p$ is the proton mass, $\sigma_v$ is the gaussian standard deviation of the observed velocity distribution, and the factor 1.4 accounts for helium.  Using Parkes Galactic All Sky Survey \citep{kal10} \hi\ observations towards DCLd 300.2--16.9, we measure $\sigma_v\approx 3\kms$.  Dividing the turbulent energy by the cooling rate gives the dissipation time:
\begin{align}
t_{\litl dis} & = \frac{E_{\litl kin,H}}{\Lambda_{\rm H}}\nonumber
\\            & = \nexpo{3}{5}\yr\,\left(\frac{\sigma_v}{3\kms}\right)^2\,
\left(\frac{\nexpo{5}{-9}\wmsr}{I(0-2)}\right)\nonumber
\\            & \times \left(\frac{\nnh}{\nexpo{2}{21}\cmtwo}\right).
\label{tdis}
\end{align}
Thus, the time over which the observed \htwo\ emission can dissipate the available turbulent kinetic energy is of the same order of magnitude as the time since supernova impact.  

This dissipation time might be an upper limit, since the \htwo\ lines are not the only coolant of the warm gas.  Above about 200\K\ the 63\micron\ fine-structure line of neutral oxygen [\oi ] can account for a significant fraction of the cooling; and below 200\K\ the 158\micron\ line of ionized carbon [\cii ] can dominate the cooling \citep{fal07}.  Our non-LTE+PDR model predicts 63\micron\ [\oi ] and 158\micron\ [\cii ] emission from the $T=391\K$ warm zone of $\approx\nexpo{1}{-9}\wmsr$ each, which increases the total cooling by about 40\% over that due to \htwo\ emission alone, leading to a corresponding decrease in the estimated dissipation time.  (The model assumes an oxygen abundance of $\nexpo{4}{-4}$ and a carbon abundance of $\nexpo{1}{-4}$.)  For DCld 300.2--16.9, the reduced dissipation time is about the same as the time since supernova impact deduced by \citet{neh08a} ($\nexpo{2-3}{5}\yr$, based on the time it took the observed 10\kms\ \hi\ streaming motions to create the cloud's 2.4\pc\ tail).  The detection of [\oi ] 63\micron\ emission at the $\sim\expo{-9}\wmsr$ level would help to confirm our assertion that the \htwo\ emission we observe originates in anomalously warm regions, since the model predicts about 6 times as much [\oi ] flux in the warm zone as in the PDR.  Nearly all of the HLC positions in our study show [\cii ] 158\micron\ emission \citep{ing02} at a level consistent with the PDR part of the model.  Since only 5\% of [\cii ] emission is expected to originate in the 391\K\ zone, this cannot likewise be used to corroborate the existence of warm material.

For LDN 1780, we have inferred a time of $\nexpo{6}{5}\yr$ since supernova impact, which is somewhat larger than the estimated dissipation time.  We lack S(0) or S(1) measurements towards this cloud, however, so it is not clear that $t_{\litl dis}$ estimated from the DCld 300.2--16.9 total S(0-2) intensity is appropriate for LDN 1780.  

The timescale to expend the available kinetic energy can also {\it increase} by a factor of a few or more, depending on the specific mechanism of dissipation.  If the \htwo\ emission is primarily the result of large-scale shocks, then dissipation will begin soon after energy injection.   If the emission is caused by turbulent dissipation at small scales, however, then any large-scale mechanical energy input will take additional time to flow from the input scale to the dissipation scale.  For a Kolmogorov-like turbulent spectrum the dissipationless cascade will be dominated by the large-scale turnover time, $L/\sigma_v$.  Given a velocity dispersion of 3\kms\ and an input size scale of $L\sim 1\pc$, this turnover time is $\approx\nexpo{3}{5}\yr$, of order the dissipation time itself.

In general, a given spot in the Galactic disk experiences a supernova explosion within 100\pc\ once every $\sim\expo{6}\yr$.  Since any energy deposited would dissipate in $t_{\litl dis}\approx \nexpo{3}{5}\yr$, our scenario provides a natural explanation for the statistical dearth of rotationally-excited \htwo\ in other translucent regions in our 100\pc\ neighborhood (we only detected 12.3\micron\ emission towards 2 out of 6 HLCs observed).  The case of LDN 183, considered part of the same molecular complex as LDN 1780 \citep{lau95,lal03}, lends plausibility to this picture. LDN 183, which lacks detectable S(2) emission, is {\it outside} the arc of the \hi\ loop that passed through LDN 1780 \citep{tot95} $\sim 6\times\expo{5}\,$yr ago, and so perhaps has not yet been subjected to the same large scale energy injection that excited the S(2) emission we see towards LDN 1780.

\subsection{A New Source of \htwo\ Photons from Galaxies?}
DCld 300.2--16.9 and LDN 1780 could be templates for an as-yet undefined ``translucent''  source of \htwo\ pure-rotational emission from the disks of all non-active galaxies.  For a large sample of galaxies, \citet{rou07} found that the ratio between the total \htwo\ flux in the S(0)--S(2) transitions and the PAH emission at 7.9\micron, F(S0-S2)/F$_{\litl 7.9,PAH}$, was remarkably constant.  The ratio was constant regardless of whether the IR emission was dominated by cirrus or high-intensity PDRs (their Figure 9), which they took to signify that the emission from both PAHs and \htwo\ is caused chiefly by FUV excitation in PDRs.  But when we compare our \Spitzer\ F(S2)/F$_{\litl 7.9,PAH}$ measurements to the \citeauthor{rou07} data (Figure \ref{hlc_roussel_h2}), we see that cirrus positions are paradoxically {\it brighter} in S(2) emission (relative to PAH emission), by a factor of 4 (on average).  Thus translucent material appears to be at least as effective at producing rotationally excited \htwo\ molecules (per PAH-exciting photon) as PDRs surrounding \hii\ regions.

Including the S(0) and S(1) emission for DCld 300.2--16.9 and estimating the TIR emission yields S(0--2)/TIR values that are $\sim 10$ times higher on average than the non-active galaxy ensemble.  This implies that the overall \htwo\ excitation in translucent cirrus is higher per unit of {\it total} energy absorbed by dust from the ambient radiation field.  Qualitatively, this is not a controversial statement; \citet{bla87} showed that regions with $G_0/n\ll 1$ are more efficient at radiatively exciting \htwo\ than regions with $G_0/n \gtrsim 1$, because when $G_0/n$ is high the resulting low \htwo\ abundance means that the 1110--912\AA\ UV photons that might pump \htwo\ are instead absorbed by dust grains.  But as we have shown, the S(0) to S(2) emission we observe is far greater than that expected from PDR models, requiring us to invoke collisional excitation in anomalously warm regions.

The HLCs in which we have detected S(2) emission have a greater concentration of rotationally warm gas than the integrated disks of galaxies---including the Milky Way.  The integrated S(2) emission from a column of material equivalent to $\sim 30\,$mag of visual extinction in the Milky Way \citep{fal05} is indistinguishable on Figure \ref{hlc_roussel_h2} from the SINGS star-forming galaxies.  One-half of the mass in the MW column is calculated to be in diffuse gas \citep{fal05}, so it is reasonable to expect that other galaxies likewise have a significant mass of diffuse gas that emits S(2) radiation. Indeed, there is evidence that cold gas with detectable S(2) emission, if not common, may nevertheless be present throughout the disks of normal galaxies.  As mentioned, the data are consistent with other measurements of nearby molecular lines of sight:  analysis of both the emission from Taurus \citep{gol10} and local absorption towards stars measured with {\sl FUSE} \citep{rac01,gry02} shows similar excitation conditions to those reported here (\S3.2).  In addition, recent measurements of \htwo\ in nearby galaxies \citep{lai10,sta10} yield S(0)/S(1) rotational temperatures that are similar to those we measure ($\sim 100-150\K$).  These temperatures do not vary across the disks of the galaxies, despite the fact that star formation indicators decrease radially, suggesting a component of excitation that is unrelated to PDRs surrounding O and B stars.

Assume that the 34-position high--latitude cloud sample is representative of all translucent gas in the Galaxy.  If so, then the average F(S2)/F$_{\litl 7.9,PAH}$ ratio for all positions, including nondetections, estimates the overall Galactic ratio for such material.  A gray dashed ellipse on Figure \ref{hlc_roussel_h2} represents this average, and is centered on practically the same value of F(S2)/F$_{\litl 7.9,PAH}$ as the SINGS average.  It is therefore plausible that translucent clouds integrated over the disks of galaxies represent a significant source of the \htwo\ pure-rotational emission.

\subsection{Future Work to Characterize Warm \htwo\ in Cirrus Clouds}
We have proposed a rather speculative scenario to explain the S(0) through S(2) emission we detect towards two translucent cirrus clouds.  We used models of radiative and mechanical heating to estimate the conditions that could reproduce the observations.  Knowledge of the superbubble environment of the clouds lends credence to the idea that extra mechanical energy is available to raise enough \htwo\ to the $J=4$ state.  The processes responsible for the heating remain uncertain.  Turbulent dissipation regions \citep[TDRs;][]{god09} may be able to produce strong localized heating, but more complete theoretical models of how this may proceed in magnetized interstellar gas remain to be developed.  If shock waves are responsible for the heating, it should be possible to identify other evidence for the presence of a shock, such as separate radial velocities in the preshock and postshock regions.

New observations are also required to confirm our interpretation of the data, especially our assertion that \htwo\ rotational emission from cirrus is widespread in normal galaxies, and that cirrus gas is a major source of the integrated emission.  \htwo\ absorption line studies with {\sl FUSE} have already established the ubiquity of \htwo\ in $J=2$, 3, 4, and 5 at levels in excess of what can be explained by UV pumping \citep{wak06,gil06}---see Figure 31.4 in \citet{dra11}.  The Cosmic Origins Spectrograph on Hubble Space Telescope can make additional \htwo\ absorption line measurements in diffuse and translucent clouds when suitable background UV sources are available.  Later in this decade, it will be possible to measure the IR line emission from rotationally-excited \htwo\ using both the James Webb Space Telescope (JWST) and the proposed Space Infrared Telescope for Cosmology and Astrophysics (SPICA).  Other tracers of warm diffuse molecular gas, such as CH$^+$ and SH$^+$ \citep[e.g.,][]{men11,fal10a,fal10b}, and [\oi], can also be used to study the heating processes in diffuse and translucent material.

\section{Conclusions}

We have detected molecular hydrogen pure-rotational ($v=0-0$) emission up to the S(2) transition towards 6 positions in two nearby high Galactic latitude translucent cirrus clouds (HLCs).  Excitation analysis  for the three available lines show that a small amount ($\lesssim 2\%$) of the gas is rotationally warm ($\trot \gtrsim 300\K$).  Photodissociation region models with $G_0 = 0.6-10$ are unable to account for the observed emission, whereas models of shocks or turbulent dissipation regions can reproduce the S(1) and S(2) measurements. Both clouds have likely been impacted by shock waves from the Scorpius-Centaurus OB association about $2-6\times\expo{5}\yr$ ago, which may have injected sufficient mechanical energy to excite the \htwo\ rotational levels.  Four other nearby HLCs, which are not so situated, did not show detectable S(2) lines.  This picture is still speculative, however, and requires robust modeling to confirm it.

Prior to the \Spitzer\ Space Telescope, published observations of molecular hydrogen in the diffuse interstellar medium were confined to UV absorption measurements \citep[for example, see][]{rac01,rac02,gry02,gil06,wak06,rac09,jen10} or a measurement of the emission along an extensive ($\av\sim 30$\,mag) line of sight in the Galactic midplane \citep{fal05}.  During its cryogenic lifetime, \Spitzer\ has enabled a new way of looking at the neutral ISM, confirming the notion that a significant amount of warm molecular gas exists within the cold medium, both within our own Galaxy \citep[this paper;][]{gol10}, and in external galaxies \citep{lai10,sta10}.  We propose that rotationally warm gas inside cold clouds is a feature of the disks of all normal galaxies.  While it may not be common (less than 20\% of our sample had detectable \htwo\ emission, and at most 2\% of the column density towards the detected positions is warm gas), it appears to be widespread.  The enhanced \htwo\ flux relative to dust emission in HLCs compared with the inner disks of the SINGS galaxies, together with the known ubiquity of $\av\sim 1$ material in the Galactic plane \citep{pol88}, requires that the translucent cirrus be taken into account when interpreting the \htwo\ rotational lines in the disks of all non-active galaxies.

\acknowledgements
This work is based on observations made with the \Spitzer\ Space Telescope, which is operated by the Jet Propulsion Laboratory, California Institute of Technology under a contract with NASA. Support for this work was provided by NASA through an award issued by JPL/Caltech.  We thank the anonymous referee for comments which improved this paper. {\it Facilities:} \facility{Spitzer ()}

\end{document}